\documentclass[usenatbib]{mn2e}
\def\kms{\,{\rm km}\,{\rm s}^{-1}}

\usepackage{epsfig,colordvi}

\usepackage{amssymb}
\usepackage{amsmath}

\newcommand{\mnras}{Mon. Not. R. ast. Soc.}
\newcommand{\nilc}{{\tt NILC\,}}
\newcommand{\sevem}{{\tt SEVEM\,}}
\newcommand{\smica}{{\tt SMICA\,}}
\newcommand{\commander}{{\tt COMMANDER\,}}

\def \der{{\rm d}}
\def \me{{\mathrm{e}}}
\def\kms{\,{\rm km}\,{\rm s}^{-1}}
\def \msun{{{\rm M}_{\odot}}}
\def \r1r2{{|{\bf r}_{1}-{\bf r}_{2}|}}

\def\mnras{Monthly Notices of the Royal Astronomical Society}
\def\apj{The Astrophysical Journal}

\def\aap{Astronomy and Astrophysics}
\def\physrep{Physics Reports}
\def\jcap{Journal of Cosmology and Astroparticle Physics}
\def\prd{Physical Review D}

\def\aj{The Astronomical Journal}

\begin{document}

\title[Velocity bias and optical depth]{Constraining the optical depth of galaxies and velocity bias with cross-correlation between kinetic Sunyaev-Zeldovich effect and peculiar velocity field}

\author[Ma, Gong, Sui, \& He]{Yin-Zhe Ma$^{1,2,\dagger}$, Guo-Dong Gong$^{3,2}$, Ning Sui$^{3,2}$, \& Ping He$^{3,4,2\star}$ \\
$^1$School of Chemistry and Physics, University of KwaZulu-Natal, Westville Campus, Private Bag X54001, Durban, 4000, South Africa.\\
$^2$NAOC-–UKZN Computational Astrophysics Centre (NUCAC), University of KwaZulu-Natal, Durban, 4000, South Africa.\\
$^{3}$College of Physics, Jilin University, Changchun 130012, PR China. \\
$^{4}$Center for High Energy Physics, Peking University, Beijing 100871, PR China. \\
emails: $^{\dagger}$ma@ukzn.ac.za;\,
$^{\star}$hep@itp.ac.cn}

%

\maketitle

\begin{abstract}
We calculate the cross-correlation function $\langle (\Delta T/T)(\mathbf{v}\cdot \hat{\mathbf{n}}/\sigma_{v}) \rangle$ between the kinetic Sunyaev-Zeldovich (kSZ) effect and the reconstructed peculiar velocity field using linear perturbation theory, with the aim of constraining the optical depth $\tau$ and peculiar velocity bias of central galaxies with {\it Planck} data. We vary the optical depth $\tau$ and the velocity bias function $b_{v}(k)=1+b(k/k_{0})^{n}$, and fit the model to the data, with and without varying the calibration parameter $y_{0}$ that controls the vertical shift of the correlation function. By constructing a likelihood function and constraining $\tau$, $b$ and $n$ parameters, we find that the quadratic power-law model of velocity bias $b_{v}(k)=1+b(k/k_{0})^{2}$ provides the best-fit to the data. The best-fit values are $\tau=(1.18 \pm 0.24) \times 10^{-4}$, $b=-0.84^{+0.16}_{-0.20}$ and $y_{0}=(12.39^{+3.65}_{-3.66})\times 10^{-9}$ ($68\%$ confidence level). The probability of $b>0$ is only $3.12 \times 10^{-8}$ for the parameter $b$, which clearly suggests a detection of scale-dependent velocity bias. The fitting results indicate that the large-scale ($k \leq 0.1\,h\,{\rm Mpc}^{-1}$) velocity bias is unity, while on small scales the bias tends to become negative. The value of $\tau$ is consistent with the stellar mass--halo mass and optical depth relation proposed in the previous literatures, and the negative velocity bias on small scales is consistent with the peak background-split theory. Our method provides a direct tool to study the gaseous and kinematic properties of galaxies.
\end{abstract}


\section{Introduction}
\label{sec:intro}
The kinetic Sunyaev-Zeldovich effect (kSZ, ~\citet{Sunyaev72,Sunyaev80}) describes the temperature anisotropy of the cosmic microwave background (CMB) radiation due to CMB photons scattering by clouds of electrons moving with non-zero line-of-sight velocities with respect to the CMB rest frame. It was proposed by~\citet{Sunyaev72,Sunyaev80} that
\begin{eqnarray}
\frac{\Delta T}{T}(\mathbf{\hat{n}})=-\frac{\sigma_{\rm T}}{c} \int n_{\rm e} \left(\mathbf{v}\cdot \mathbf{\hat{n}} \right) \der l ,\label{eq:kSZ1}
\end{eqnarray}
where $\sigma_{\rm T}$ is the Thomson cross-section, $n_{\rm e}$ is the electron density, $\mathbf{v}\cdot \mathbf{\hat{n}}$ is the velocity along the line-of-sight, and $\der l$ is the differential of proper distance $l$ in the radial direction. Note that this is the non-relativistic kSZ equation, which is quite accurate for coherent flow $\lesssim 500\kms$. \citet{Nozawa98} and \citet{Nozawa15} calculated the (higher-order) relativistic corrections to the kSZ effect, which is at order $\lesssim\mathcal{O}(0.8\%)$ for the two second-order corrections. Therefore the relativistic corrections are generally negligible.

By using the pairwise momentum estimator, which quantifies the difference in temperature between pairs of galaxies, the kSZ effect was first detected by \citet{Hand12} using data from the Acatama Cosmology Telescope (ACT). The detection of the kSZ effect was further solidified by the application of the same pairwise momentum estimator to other CMB data, including
the {\it Wilkinson Microwave Anisotropy Probe} ({\it WMAP}) 9-year W-band data, and four {\it Planck} foreground-cleaned maps~(\citet{Planck16-unbound}). The measurements are at $3.3\sigma$ and $1.8$--$2.5\sigma$ confidence level (C.L.)
respectively for {\it WMAP} and {\it Planck}. More recently, \citet{Hill16} cross-correlated the square field of the kSZ measured by {\it WMAP} and {\it Planck} surveys with the projected galaxy overdensity from the {\it Wide-field Infrared Survey Explorer} ({\it WISE}) and reported a $3.8\sigma$ C.L. detection. \citet{Ferraro16} forecasted that for Advanced ACTPol and hypothetical Stage-IV CMB experiment the signal-to-noise ratio can reach $120$ and $150$ respectively. \citet{Schaan16} detected the aggregated signal of kSZ at $\sim 3.3\sigma$ C.L. by cross-correlating the velocity field from CMASS samples (BOSS-DR10) with the kSZ map produced from ACT observations. \citet{Bernardis16} applied the pairwise momentum estimator to the ACT data and $50,000$ bright galaxies from the BOSS survey, and obtained a $3.6 \sigma$--$4.1\sigma$ C.L. detection. \citet{Soergel} also used the pairwise momentum estimator, applied to South Pole Telescope (SPT) data and photometric survey data from Dark Energy Survey (DES), and obtained the averaged central optical depth of galaxy clusters at $4.2\sigma$ C.L. 

In order to trace the centers of dark matter halos, \citet{Planck16-unbound} constructed a Central Galaxy Catalogue (CGC), a galaxy sample composed of $262,673$ spectroscopic sources brighter than $r = 17.7$. They applied some isolation criterion to extract these sources from Sloan Digital Sky Survey's Data Release 7 (SDSS-DR7), so that these sources, at least a large fraction of them, can be regarded as representatives of {\em isolated} or {\em field} galaxies. By using the continuity equation, \citet{Planck16-unbound} reconstructed the linear velocity field from galaxies of the CGC. With $\langle \Delta T (\mathbf{v}\cdot \mathbf{\hat{n}}) \rangle$ the estimator that cross-correlates the kSZ with the reconstructed velocity field, they made detections at $3.0$--$3.2\sigma$ C.L. using the four foreground-cleaned {\it Planck} maps, and the detection at $3.8\sigma$ C.L. using the {\it Planck} 217 GHz raw map. \citet{Planck16-unbound} approximated the kSZ equation (\ref{eq:kSZ1}) as
\begin{eqnarray}
\frac{\Delta T(\hat{\mathbf{n}})}{T} & \simeq & -\sigma_{\rm T} \left(\frac{\mathbf{v}\cdot \hat{\mathbf{n}}}{c} \right) \int \der l \, n_{\rm e} \nonumber \\
&= & - \left(\frac{\mathbf{v}\cdot \hat{\mathbf{n}}}{c} \right) \tau, \label{eq:kSZ2}
\end{eqnarray}
where $\tau=\sigma_{\rm T}\int \der l n_{\rm e}$ is the optical depth for the CGC galaxies. Equation~(\ref{eq:kSZ2}) takes the velocity out of the integral because the correlation length of bulk motion $\mathbf{v}$ is much larger than the variation in the electron density of galaxies. In addition, fitting the cross-correlated $\langle \Delta T (\mathbf{v}\cdot \mathbf{\hat{n}}) \rangle$ data with $N$-body simulation, the average optical depth is found to be $\tau=(1.39 \pm 0.46) \times 10^{-4}$ ($1\sigma$ C.L.) for \sevem map~(\citet{Planck16-unbound}). A follow-up paper~\citep{Carlos} showed that the majority of baryons are located outside the virial radii of galaxy halos.

In this paper, we use a different method to calculate the kSZ--velocity field cross-correlation, by using the linear perturbation theory to fit for the optical depth of CGC galaxies. In addition, we vary the velocity bias function to fit the kSZ--velocity field cross-correlation data. The purpose of this work is to provide complementary constraint on average optical depth $\tau$ and probe the bias of the peculiar velocity field. The peculiar velocity field bias $b_{v}$ is generally assumed to unity on large scales. However, this fundamental assumption has not been systematically tested and verified with numerical simulations. The obstacle lies in how to tackle the unphysical sampling artifact, which is entangled in the measured velocity statistics and becomes significant for sparse populations~\citep{Zhang15,Zheng15a,Zheng15b}. Recently, by using $N$-body simulations, ~\citet{Zheng15b} verified that $b_{v}=1$ within $2\%$ of the model uncertainty at $k\leq 0.1\,h\,{\rm Mpc}^{-1}$, in the redshift range $0<z<2$ for halos of mass in the range of $10^{12}$--$10^{13}\,h^{-1}M_{\odot}$; while at $k \geq 0.1\,h\,{\rm Mpc}^{-1}$, $b_{v} \neq 1$. For additional references that investigate the velocity bias on large and small scales, please refer to~\citet{Elia12,Biagetti14,Baldauf15,Chan15,Guo15a,Guo15b}. We will explore the same problem but using the cross-correlation data from kSZ and velocity field.

This paper is organized as follows. In Sec.~\ref{sec:data}, we will discuss the CMB map and galaxy survey catalogue that are used to obtain the data of kSZ--linear velocity correlation. In Sec.~\ref{sec:cross}, we present the calculation of the correlation function, the three velocity bias models under investigation in this work, and we briefly discuss the likelihood analysis. We present the results in Sec.~\ref{sec:results}, and discussion in Sec.~\ref{sec:discuss}. We make conclusion and discuss future prospects in Sec.~\ref{sec:conclusion}.

Since the typical optical depth for a galaxy is order of $\mathcal{O}(10^{-5})$--$\mathcal{O}(10^{-4})$, we define $\tau_{4}\equiv \tau \times 10^{4}$ and use it throughout the paper. In addition, we adopt a spatially flat, $\Lambda$CDM cosmology model, with the best-fit cosmological parameters given by {\it Planck} 2015 results~\citep{Planck15-para}: $\Omega_{\rm m}=0.309$; $\Omega_{\Lambda} = 0.691$; $n_{\rm s} = 0.9608$; $\sigma_{8} = 0.809$; and $h = 0.68$, where the Hubble constant is $H_{0} = 100 h \kms\,{\rm Mpc}^{-1}$.

\section{Data}
\label{sec:data}
\begin{figure}
\centerline{
\includegraphics[width=3.0in]{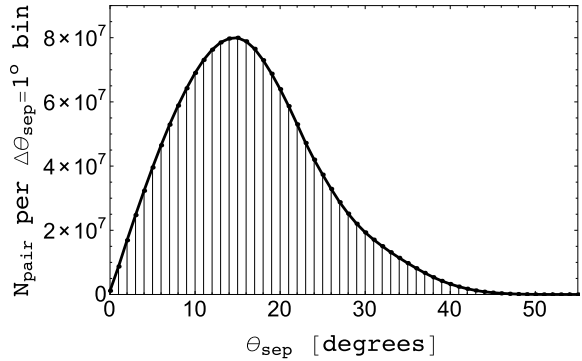}}
\caption{Histogram of the separation angles of all the pairs in the CGC, i.e. the number of pairs per $1^{\circ}$ separation bin as a function of separation angles.} \label{fig:angle}
\end{figure}

\begin{figure*}
\centerline{
\includegraphics[width=3.2in]{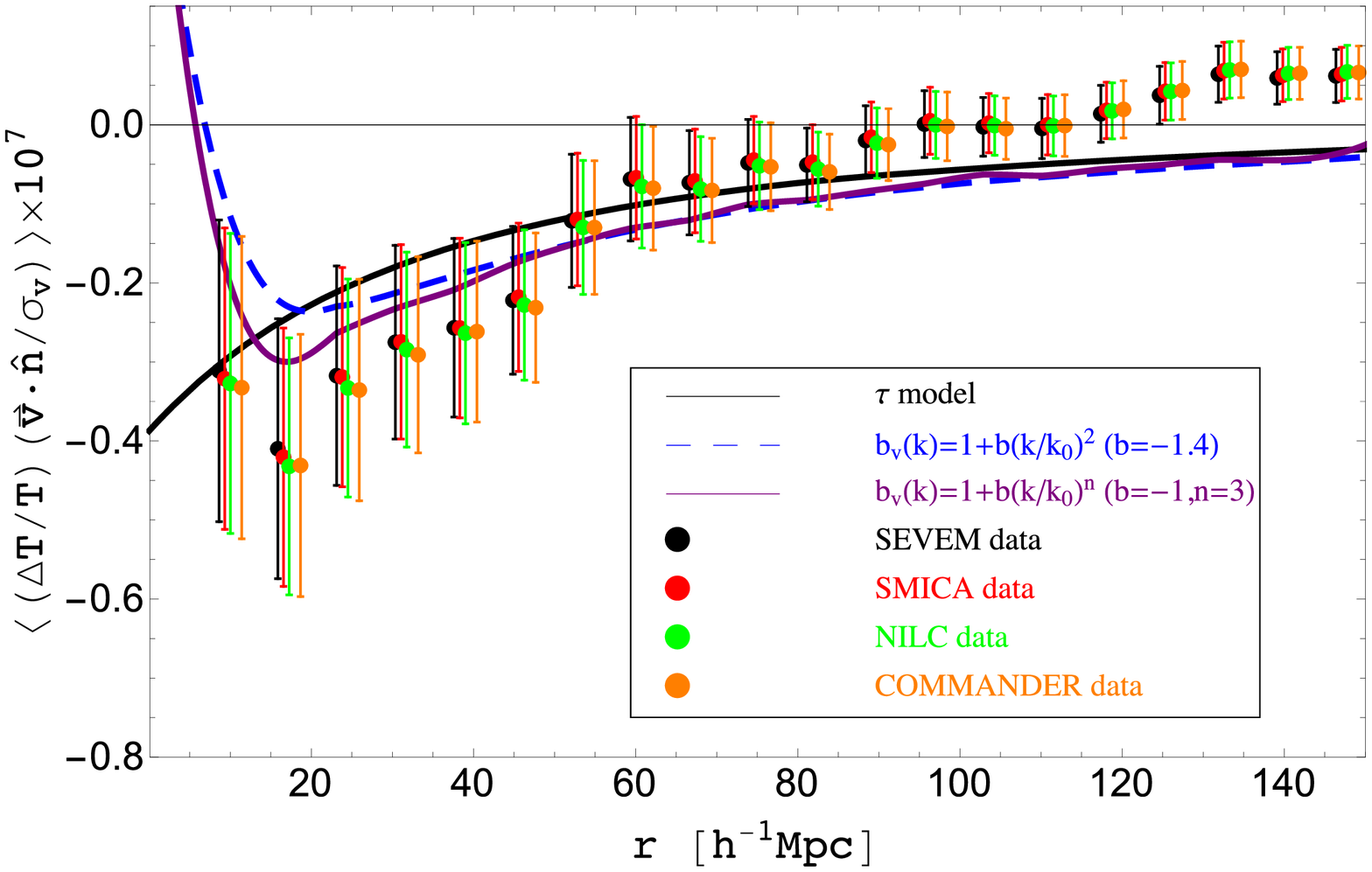}
\includegraphics[width=3.2in]{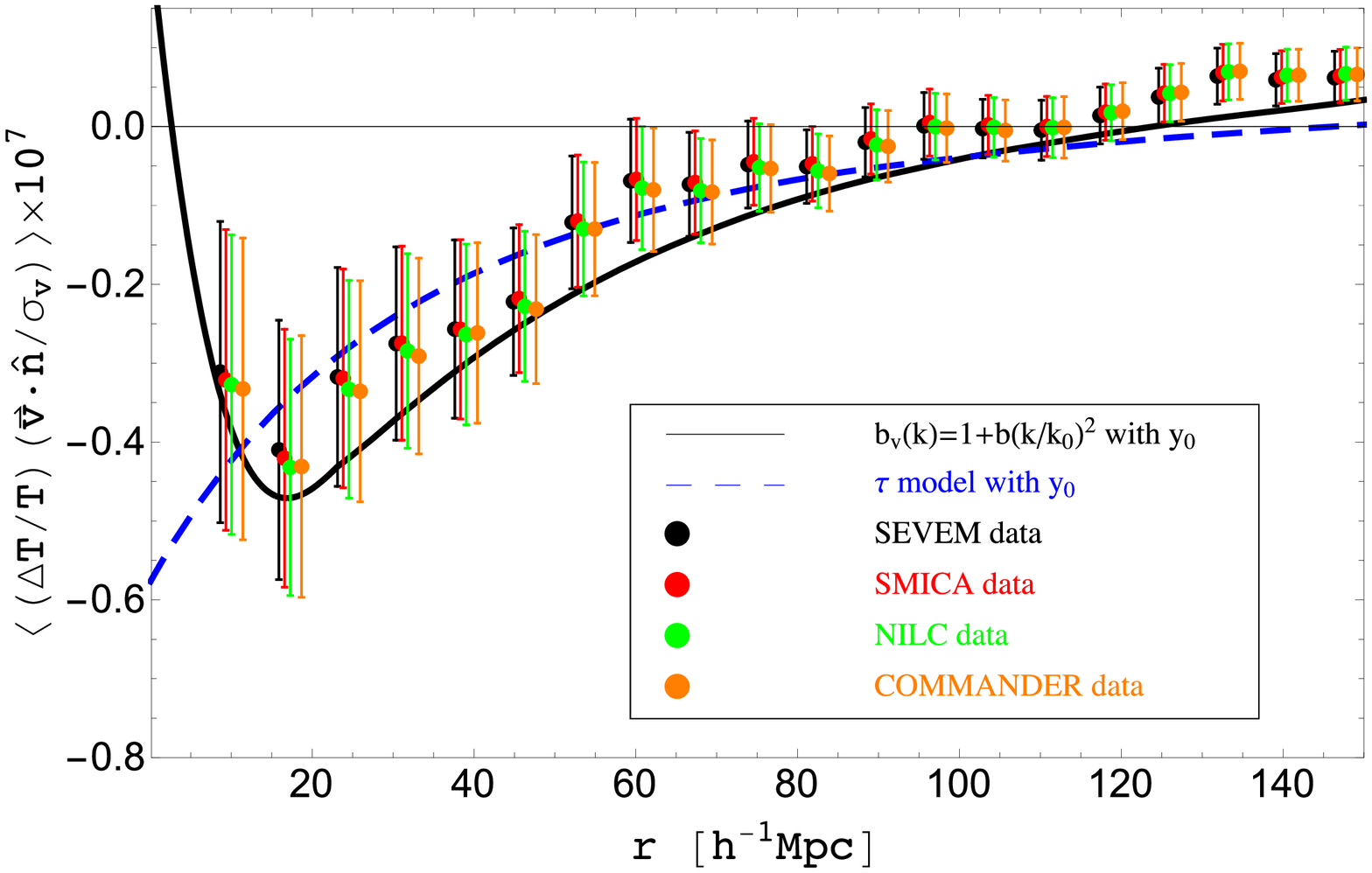}}
\caption{Comparison between data and model predictions for the kSZ--linear velocity cross-correlation. The data from {\it Planck} {\tt SEVEM}, {\tt SMICA}, {\tt NILC} and {\tt COMMANDER} maps are shown in black, red, green and organe dots. Just for visualization purpose, these data are slightly shifted horizontally to avoid overlap. Their error-bars are taken as the diagonal value of their respective covariance matrix (The correlation coefficient matrix of {\tt SEVEM} map is shown in Fig.~\ref{fig:coefffig}). {\it Left}-- The black, blue dashed and purple lines are the best-fit unbiased, quadratic power law, and varying power law models respectively without calibration parameter $y_{0}$ (Sec.~\ref{sec:bias}), i.e. for $\tau$ model, $\tau_{4}=0.40$ (black solid line); for $(\tau,b,n=2)$ model, $\tau_{4}=0.51$ and $b=-1.4$ (blue dashed line); and for $(\tau,b,n)$ model, $\tau_{4}=0.50$, $b=-1$, and $n=3$ (purple solid line). {\it Right}-- The black and blue dashed lines are best-fit unbiased and quadratic power law models with calibration parameter $y_{0}$ varying. The values are: $\tau_{4}=0.65$, $y_{9}=5.26$ (blue dashed line), and $\tau_{4}=1.18$, $b=-0.84$ and $y_{9}=12.39$ (black solid line).} \label{fig:deltaTv}
\end{figure*}


\begin{figure}
\centerline{
\includegraphics[bb=-10 -10 400 400, width=2.9in]{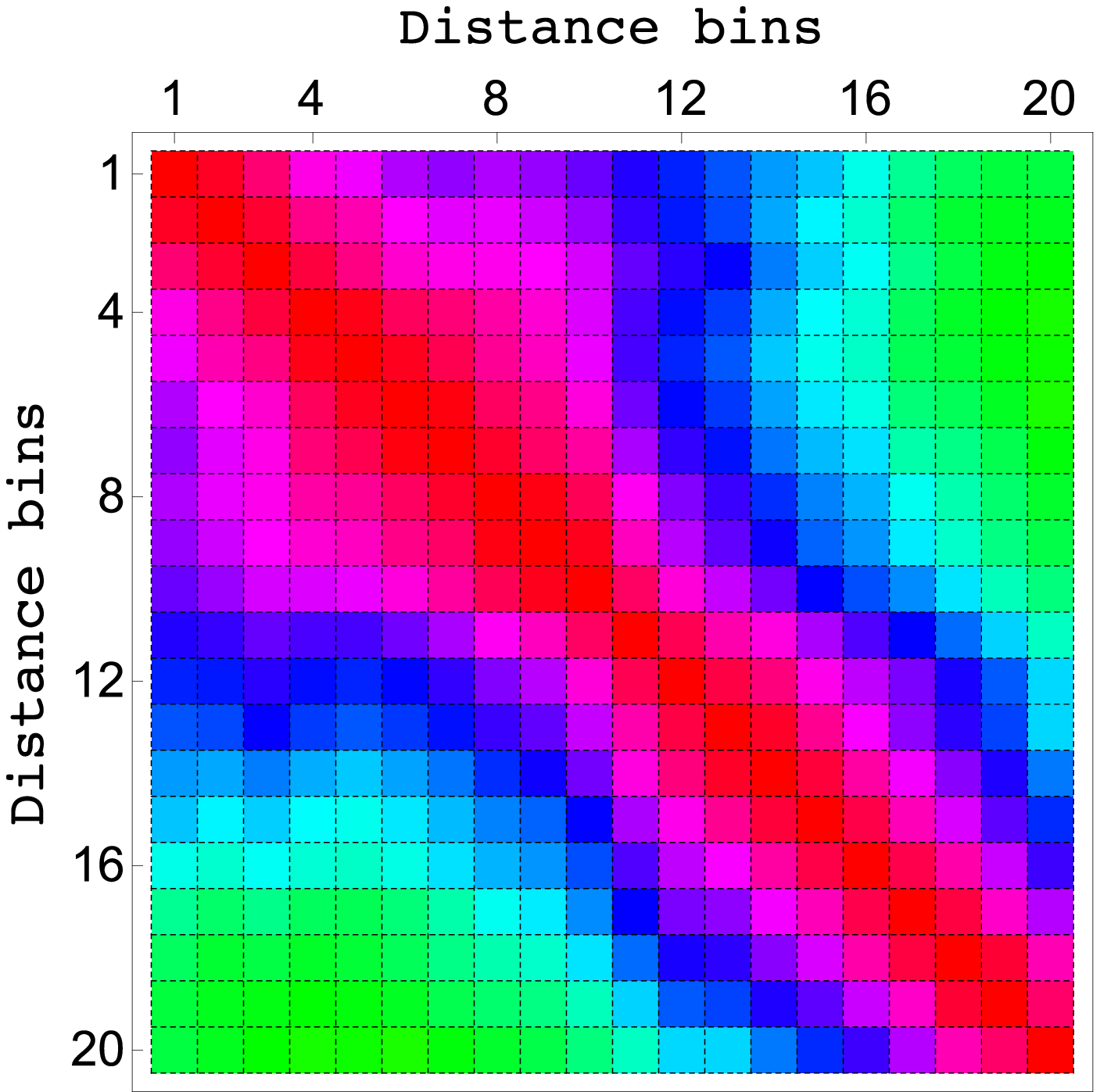}
\includegraphics[bb=-10 -10 30 30, width=0.3in]{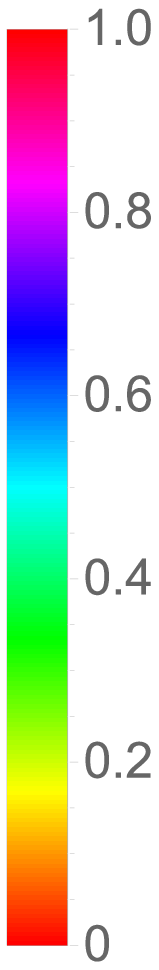}}
\caption{The correlation coefficient matrix of $\langle (\Delta T/T)(\mathbf{v}\cdot \hat{\mathbf{n}}/\sigma_{v})\rangle \times 10^{7}$ for {\tt SEVEM} map. The $x$- and $y$-axis represent distance bin index in Fig.~\ref{fig:deltaTv}.} \label{fig:coefffig}
\end{figure}


We use the results of the cross-correlation of four {\it Planck} map templates from~\citet{Planck16-unbound}. These four {\it Planck} templates are the four foreground-cleaned maps, namely the \sevem, \smica, \nilc and \commander maps. Each has a uniform beam ($\theta_{\rm FWHM}=5\,{\rm arcmin}$, ``FWHM'' stands for ``Full-Width Half Maximum''). The \smica map uses a spectral matching approach, the \sevem map uses a template-fitting method to minimize the foreground, the \nilc map is the result of an internal linear combination, and the \commander map uses a parametric, pixel based Monte Carlo Markov Chain technique to project out foregrounds~\citep{Planck15-X}. In these four maps, the thermal Sunyaev-Zeldovich (tSZ) effect, the synchrotron and dust emission are highly suppressed, while the kSZ remains in the map with the primary CMB component. This is because the spectral distortion of the kSZ is very close to the black-body radiation of CMB, and so it is not separable using only spectral filters. Using the aperture photometry method~(\citet{Planck16-unbound}) with aperture size $\theta_{\rm AP}=8\,$arcmin, the large angular temperature fluctuation is highly suppressed, and the resulting maps contain only kSZ signal, noise, the residual CMB and the tSZ effect~\citep{Planck16-unbound,Carlos}. Note that the aperture size will affect the value of the optical depth $\tau_{4}$, since the aperture size defines the typical gas cloud size measured in the kSZ observations~\citep{Planck16-unbound,Schaan16}.

In addition,~\citet{Planck16-unbound} used the Central Galaxy Catalogue (CGC) which is a data set composed of $262,673$ spectroscopic sources with stellar mass $\log(M_{\ast}/\msun)>11$, extracted from the SDSS-DR7 New York University Value Added Galaxy Catalogue~\citep{Blanton05}. After applying the criterion of brightness cut (the $r$-band extinction-corrected Petrosian magnitude, $r>17.7$), and restricting the samples within $\mathbf{R}_{\rm box}=[-300,-250,150]\,h^{-1}{\rm Mpc}$ from the observer, the resulting number of samples is $149,127$~\citep{Planck16-unbound}. In Fig.~\ref{fig:angle}, we plot the histogram of the separation angles between all the pairs of the CGC galaxies. The histogram is plotted as the number of pairs per $1$ degree bin as a function of separation angle. As one can see, the peak of the separation angle is around $15^{\circ}$. Therefore we can treat the pairs of the temperature fluctuation on the sky as extended sources, since the pair separation is much larger than the {\it Planck} beam. Then the reconstructed three-dimensional linear velocity field $\mathbf{v}$ was calculated by using the linear continuity equation ($\dot{\delta}+\nabla \cdot \mathbf{v}=0$). This velocity field represents long-wavelength perturbation that is predicted by linear perturbation theory~\citep{Peebles93}.

In~\citet{Planck16-unbound}, the cross-correlation between kSZ temperature fluctuations and the line-of-sight velocity field was calculated as $\langle \Delta T (\mathbf{v}\cdot \hat{\mathbf{n}}/\sigma_{v}) \rangle$, where $\sigma_{v}=310\,{\rm km}\,{\rm s}^{-1}$ is the mean velocity dispersion. In this paper, we normalize the correlation function with the present CMB temperature $T=2.725\,$K to make $\langle (\Delta T/T) (\mathbf{v}\cdot \hat{\mathbf{n}}/\sigma_{v}) \rangle$ a dimensionless function. Due to its low value, we multiply this function by $10^{7}$ to make it of order unity. In~\citet{Planck16-unbound}, the covariance matrix was calculated from the {\it Planck} maps after estimating the null positions where no kSZ effect is expected. Basically, the kSZ map was randomly rotated to arbitrary angles many times, and the covariance matrix was computed as the correlation between different distance bins. Therefore, this covariance matrix automatically includes the effect of cosmic variance, which is the intrinsic variation of the kSZ effect on the sky. We directly use the covariance matrix computed from~\citet{Planck16-unbound}, so one can refer to Sec.~3.3 of~\citet{Planck16-unbound} for details of this calculation.

Figure~\ref{fig:deltaTv} plots the \sevem, \smica, \nilc and \commander data as black, red, green and orange dots, and their error-bars are taken as the diagonal value of the covariance matrix (Fig.~\ref{fig:coefffig} is the correlation coefficient matrix). One can see there is strong correlations in the range of separation distance of $20$--$30\,h^{-1}$Mpc, and gradually diminished as the separation distance increases to above $80\,h^{-1}$Mpc.


One can see that on very large scales ($r>100\,h^{-1}$Mpc), the correlation function becomes slightly positive. This can be due to two reasons. One is that on large scales cosmic variance is very significant. The other reason is that in the process of producing CMB suppressed kSZ map, the aperture photometry method was used. Therefore, at each point of CGC galaxy sample, the fluctuation within $\theta_{\rm AP}=8\,$arcmin was removed to suppress the CMB contribution, which could induce a shift of the total kSZ amplitude of the resulted map. Thus, in the likelihood analysis, we will introduce a ``calibration parameter'' $y_{0}$ to take into account the uncertainty of the total amplitude of the correlation function.

\section{kSZ--velocity field cross-correlation}
\label{sec:cross}

\begin{figure}
\centerline{
\includegraphics[width=3.4in]{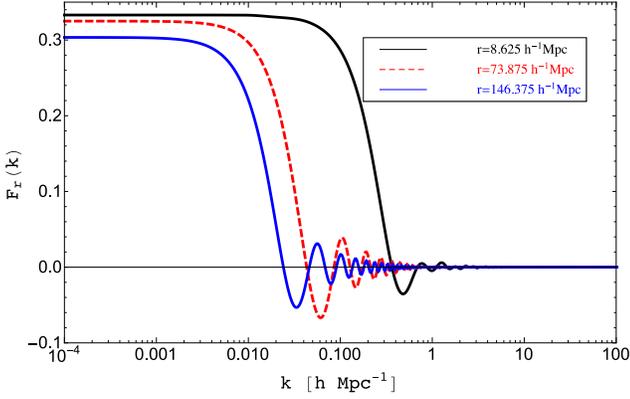}}
\caption{The $F_{r}(k)$ function for three represented distance bins centred at $r=8.625\,h^{-1}$Mpc, $r=73.875\,h^{-1}$Mpc, and $r=146.375\,h^{-1}$Mpc respectively ($\Delta r=8.725h^{-1}$Mpc). One can see that on large scales the function $F_{r}(k)$ all approach a constant value while on small scales they all fluctuate and approach zero. The larger the separation $r$ is, the quicker it falls off at large $k$.} \label{fig:Frk}
\end{figure}

\begin{figure}
\centerline{
\includegraphics[width=3.3in]{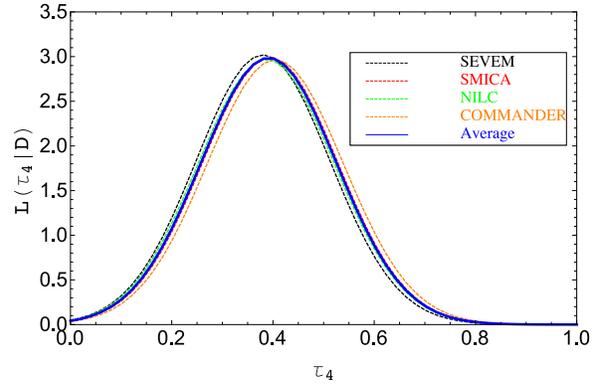}}
\caption{The likelihood function for $\tau$ parameter in the unbiased model, where only $\tau$ parameter is varied as the amplitude of correlation function. The black dashed, red dashed, green dashed and orange dashed lines are for {\tt SEVEM}, {\tt SMICA}, {\tt NILC}, {\tt COMMANDER} maps respectively. The blue solid line is the average $\chi^{2}$ result.} \label{fig:tau_1p}
\end{figure}

\begin{table*}
\begin{centering}
\begin{tabular}{|c|c|c|c|c|c|c|c|}
\hline
\noalign{\vskip 1pt}
Models & Parameters &  {\tt SEVEM} & {\tt SMICA} & {\tt NILC} & {\tt COMMANDER} & Average & $P(b>0)$ \\
\hline
$\tau$ & $\tau_{4}$ & $0.38 \pm 0.13$ & $0.39 \pm 0.13$ & $0.39 \pm 0.13$ & $0.40 \pm 0.13$ & $0.39 \pm 0.13$ & N/A  \\ \hline
$(\tau,b,n=2)$ & $\tau_{4}$ & $0.50 \pm 0.14$ & $0.50 \pm 0.14$ & $0.50^{+0.14}_{-0.15}$ & $0.52^{+0.14}_{-0.15}$ & $0.51^{+0.14}_{-0.15}$ & N/A \\ \cline{2-8} \noalign{\vskip 1pt}
 & $b$ & $-1.44^{+0.40}_{-0.58}$ & $-1.44^{+0.41}_{-0.60}$ & $-1.41^{+0.40}_{-0.58}$ & $-1.30^{+0.37}_{-0.52}$ & $-1.39^{+0.40}_{-0.57}$ & $4.47 \times 10^{-6}$ \\ \hline
 & $\tau_{4}$ & $0.50^{+0.14}_{-0.15}$ & $0.50 \pm 0.15$ & $0.50 \pm 0.15$ & $0.52 \pm 0.15$ & $0.50 \pm 0.15$ & N/A \\ \cline{2-8} \noalign{\vskip 1pt}
$(\tau,b,n)$ & $b$ & $-1.11^{+0.37}_{-0.55}$ & $-1.11^{+0.38}_{-0.56}$ & $-1.03^{+0.36}_{-0.54}$ & $-0.91^{+0.32}_{-0.48}$ & $-1.04^{+0.36}_{-0.53}$ & $6.46 \times 10^{-6}$ \\ \cline{2-8}  \noalign{\vskip 1pt}
 & $n$ & $2.94^{+0.21}_{-0.53}$ & $2.95^{+0.21}_{-0.48}$ & $3.04^{+0.19}_{-0.44}$ & $3.05^{+0.19}_{-0.49}$ & $3.00^{+0.20}_{-0.49}$  & N/A \\ \hline
$(\tau,y_{0})$ & $\tau_{4}$ & $0.56\pm 0.22$ & $0.64 \pm 0.21$ & $0.68 \pm 0.21$ & $0.71 \pm 0.21$ & $0.65 \pm 0.21$ & N/A \\ \cline{2-8}
& $y_{9}$  & $3.63 \pm 3.43$ & $4.94 \pm 3.29$ & $6.06\pm 3.29$ & $6.47 \pm 3.31$ & $5.26\pm 3.31$ & N/A \\ \hline
$(\tau,b,y_{0})$ & $\tau_{4}$ & $1.18 \pm 0.25$  & $1.17 \pm 0.24$ & $1.18 \pm 0.23$ & $1.21 \pm 0.24$ & $1.18 \pm 0.24$ & N/A \\ \cline{2-8}
$(n=2)$ & $b$  & $-0.86^{+0.16}_{-0.21}$ & $-0.87^{+0.16}_{-0.21}$ & $-0.85^{+0.16}_{-0.20}$ & $-0.78^{+0.15}_{-0.18}$ & $-0.84^{+0.16}_{-0.20}$ & $3.12 \times 10^{-8}$ \\ \cline{2-8}
& $y_{9}$  & $12.08^{+3.84}_{-3.85}$ & $12.15^{+3.61}_{-3.63}$ & $12.55^{+3.57}_{-3.58}$ & $12.79^{+3.61}_{-3.62}$ & $12.39^{+3.65}_{-3.66}$ & N/A \\ \hline
$(\tau,b,n,y_{0})$ & $\tau_{4}$ & $1.16^{+0.44}_{-0.34}$ & $1.13^{+0.40}_{-0.32}$ & $1.13^{+0.41}_{-0.32}$ & $1.19^{+0.43}_{-0.33}$ & $1.15^{+0.42}_{-0.33}$ & N/A \\ \cline{2-8}
& $b$  & $-0.98^{+0.28}_{-0.18}$ & $-0.95^{+0.26}_{-0.20}$ & $-0.93^{+0.28}_{-0.20}$ & $-0.88^{+0.26}_{-0.20}$ & $-0.94^{+0.27}_{-0.20}$ & $1.48 \times 10^{-6}$ \\ \cline{2-8}
& $y_{9}$  & $12.15^{+5.46}_{-4.73}$ & $11.97^{+5.12}_{-4.44}$ & $12.24^{+5.18}_{-4.43}$ & $12.88^{+5.28}_{-4.53}$ & $12.32^{+5.27}_{-4.54}$ & N/A \\ \cline{2-8}
& $n$ & $1.77^{+0.89}_{-0.52}$ & $1.91^{+0.82}_{-0.54}$ & $1.35<n<2.93$ & $1.73^{+0.92}_{-0.54}$ & $1.81^{+0.91}_{-0.51}$ & N/A \\ \hline
{\it Planck} result & $\tau_{4}$ & $1.39 \pm 0.46$ & N/A & N/A & N/A & N/A & N/A  \\
\hline
\end{tabular}
\caption{The best-fit parameters with $68\%$ confidence level for three different models listed in Sec.~\ref{sec:bias} ($\tau_{4}\equiv \tau \times 10^{4}$). The $y_{9}$ is defined as $y_9 \equiv y_{0}\times 10^{9}$. The last column is the probability value for $b>0$, obtained by integrating the probability function for the positive range of $b$.} \label{tab:paras}
\end{centering}
\end{table*}

\begin{table*}
\begin{centering}
\begin{tabular}{|c|c|c|c|c|c|c|}
\hline
\noalign{\vskip 1pt}
Models & parameters & {\tt SEVEM} & {\tt SMICA} & {\tt NILC} & {\tt COMMANDER} & $\overline{\chi^{2}}_{\rm min}/N_{\rm dof}$ \\
\hline
No $y_{0}$ & $\tau$ & $55.4$ & $56.6$ & $55.0$ & $54.1$ & $2.91$ \\ \cline{2-7}
calibration & $(\tau,b,n=2)$ & $33.0$ & $34.9$ & $35.0$ & $35.9$ & $1.93$  \\ \cline{2-7}
& $(\tau,b,n)$ & $29.3$ & $30.7$ & $29.6$ & $30.9$ & $1.77$  \\
\hline
with $y_{0}$ & $(\tau,y_{0})$ & $54.3$ & $54.3$ & $51.6$ & $50.5$ &  $2.93$ \\ \cline{2-7}
calibration & $(\tau,b,y_{0})$ ($n=2$) & $21.6$ & $22.1$ & $22.2$ & $21.8$ &  $1.23$  \\ \cline{2-7}
& $(\tau,b,y_{0},n)$ & $21.6$ & $22.1$ & $20.8$ & $21.8$ & $1.35$  \\ \hline
\end{tabular}
\caption{The minimal $\chi^{2}$ values from different kSZ--velocity field correlation data for the three models. The last column is the average minimal $\chi^{2}$ divided by the number of degree of freedom $N_{\rm dof}$, where $N_{\rm dof}=N_{\rm data}-N_{\rm para}$.} \label{tab:chi2}
\end{centering}
\end{table*}

\begin{figure*}
\centerline{
\includegraphics[width=3.0in]{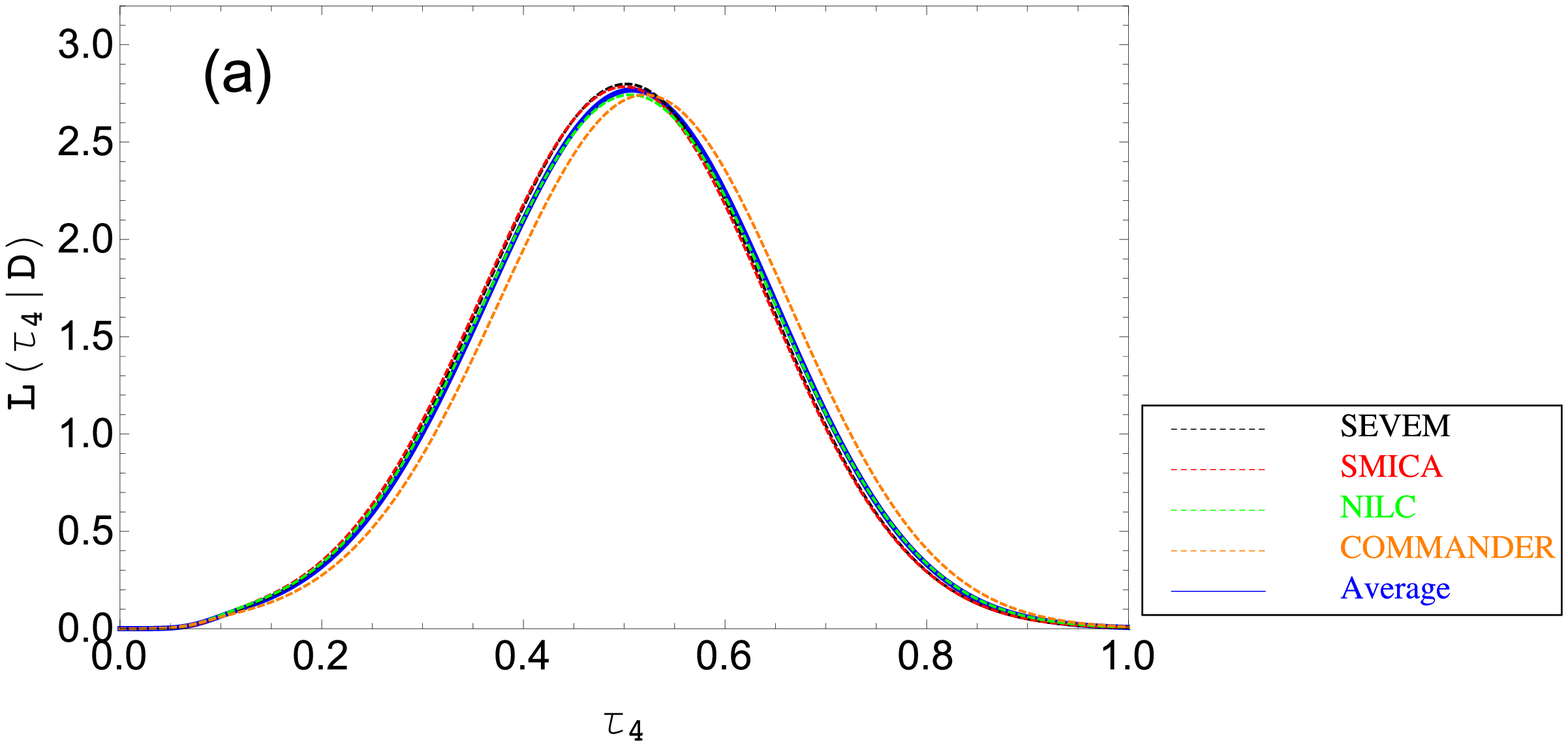}
\includegraphics[width=2.2in]{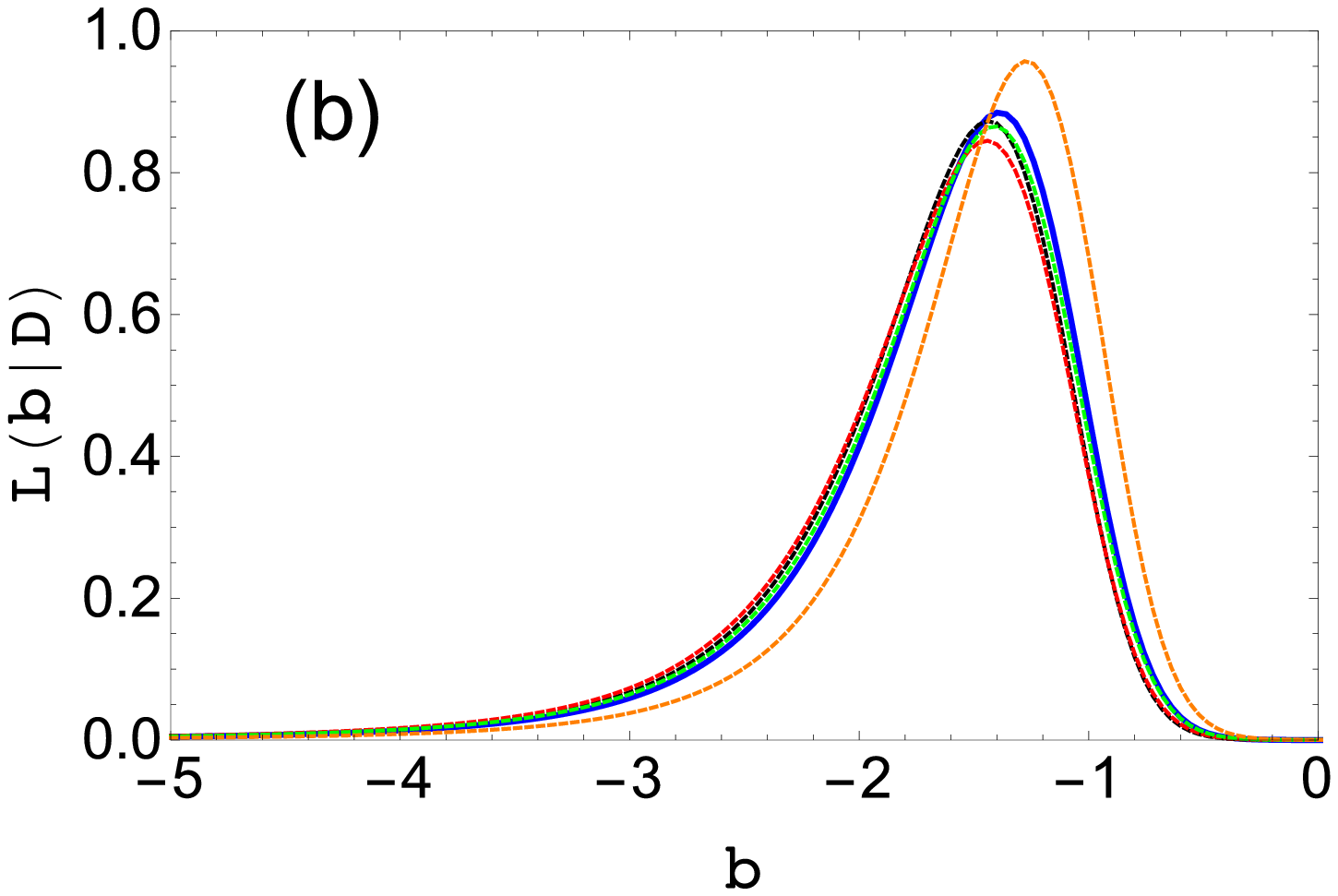}
\includegraphics[width=1.7in]{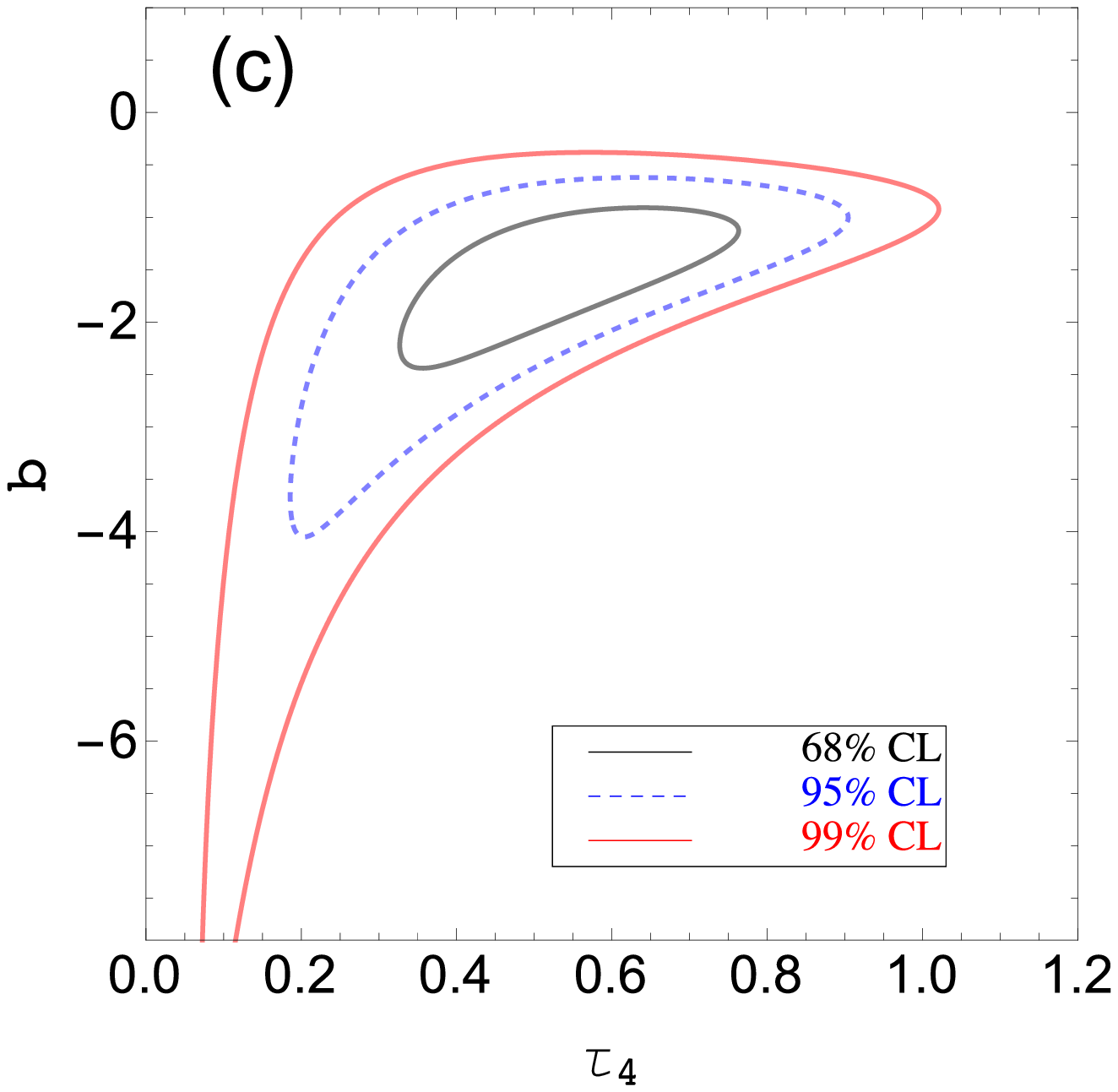}}
\caption{The fitting results for the quadratic power law model ($\tau,b,n=2$). {\it Panel (a)}: the marginalized distribution function of $\tau$; {\it Panel (b)}: the marginalized distribution function of $b$. In panels (a) and (b), the black dashed, red dashed, green dashed, and orange dashed lines are for {\tt SEVEM}, {\tt SMICA}, {\tt NILC} and {\tt COMMANDER} maps respectively. The blue solid line is for the average $\chi^{2}$ of the previous four data sets. {\it Panel (c)}: the joint constraint on $(\tau,b)$ from the averaged $\chi^{2}$ function.} \label{fig:fit_2p}
\end{figure*}

\begin{figure*}
\centerline{
\includegraphics[width=2.1in]{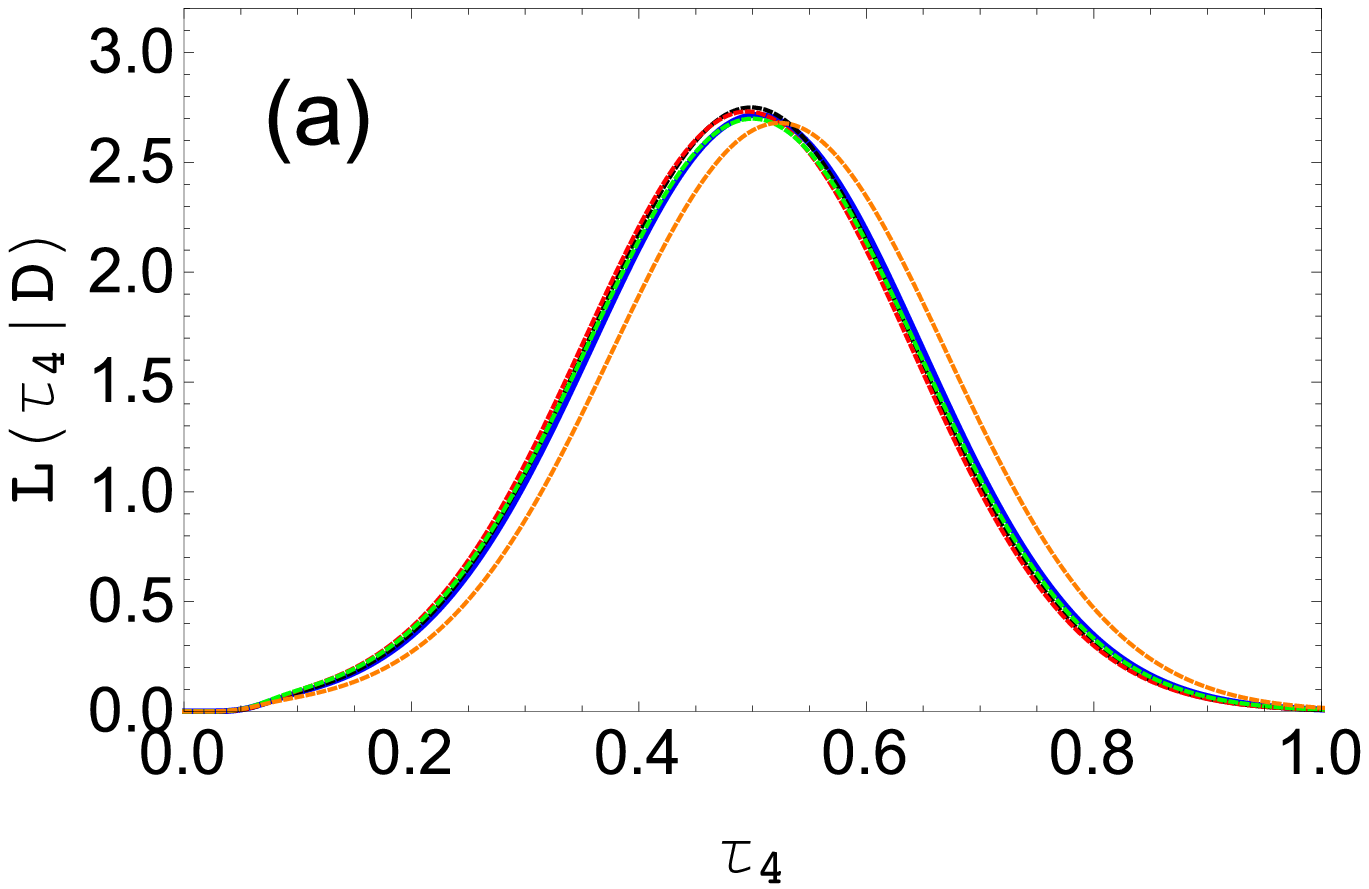}
\includegraphics[width=2.2in]{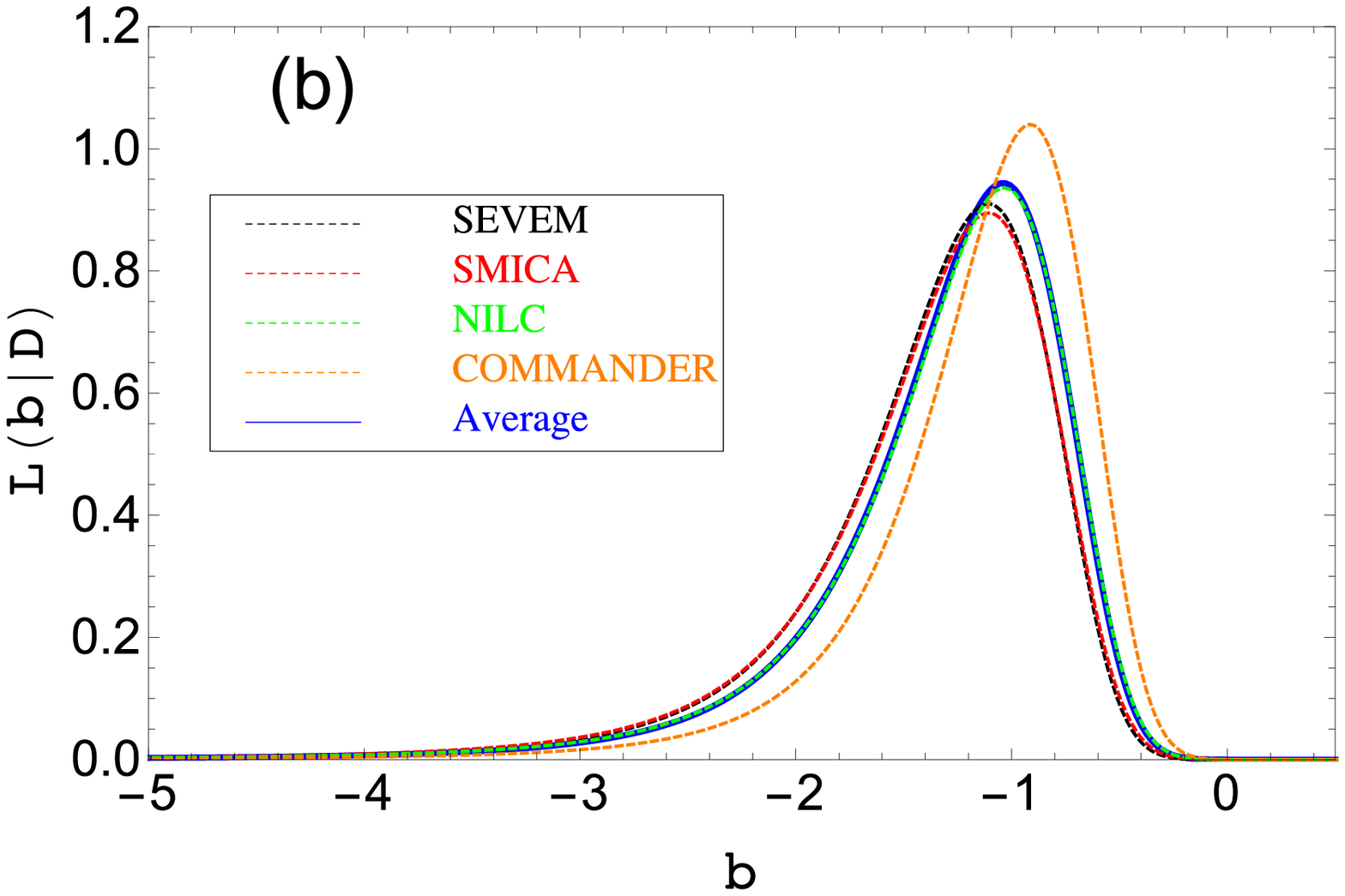}
\includegraphics[width=2.1in]{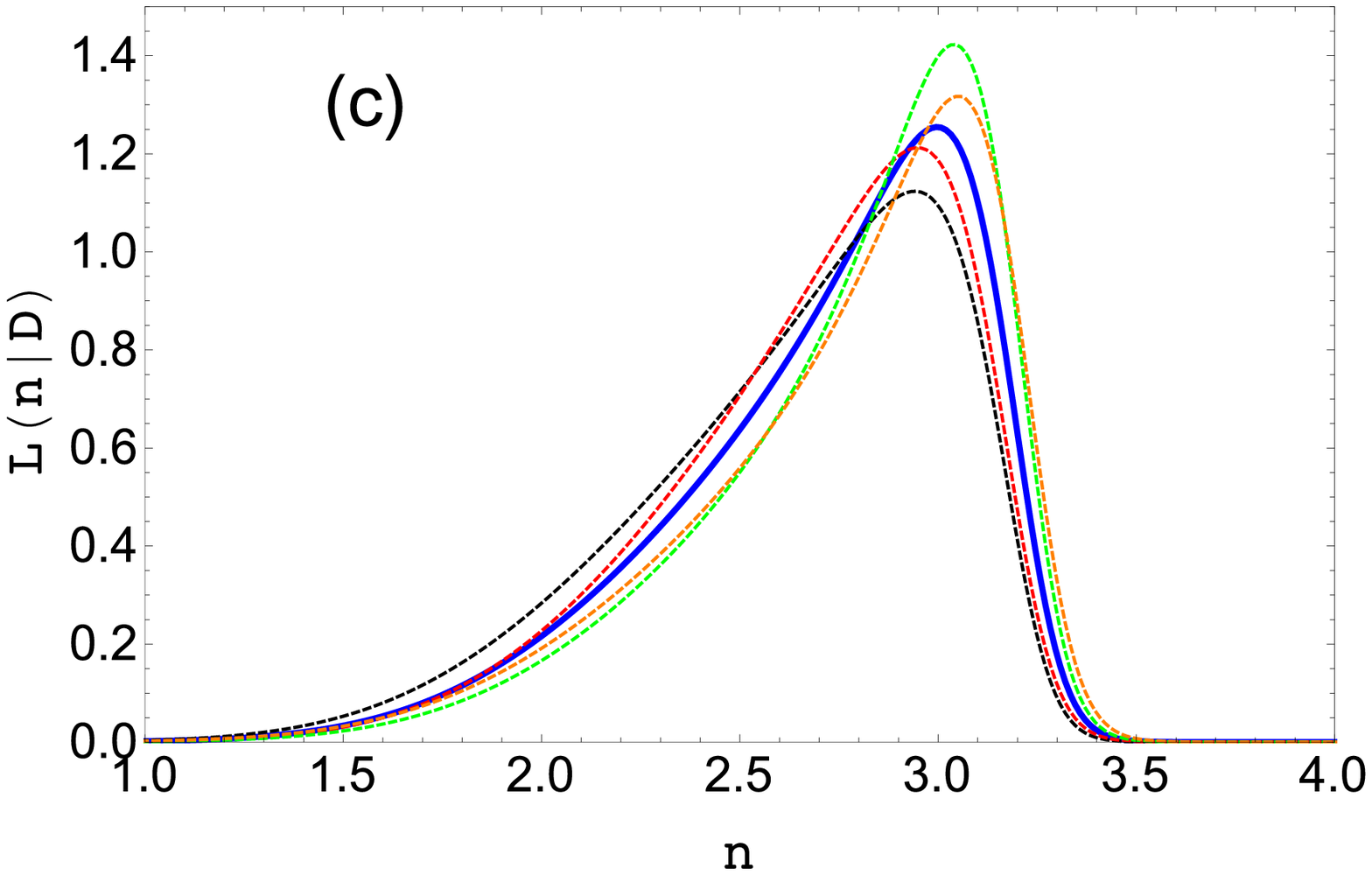}}
\centerline{
\includegraphics[width=2.2in]{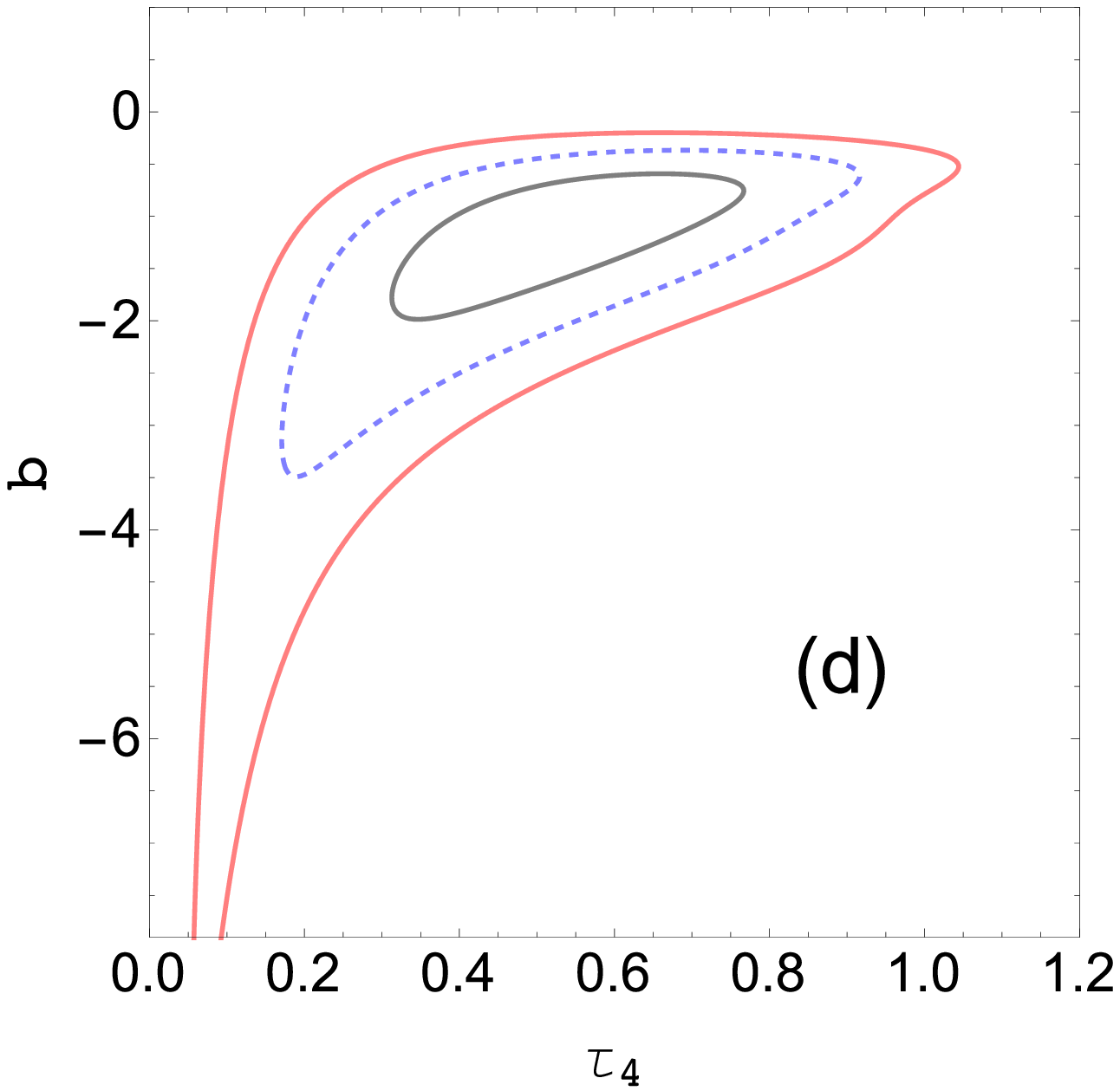}
\includegraphics[width=2.15in]{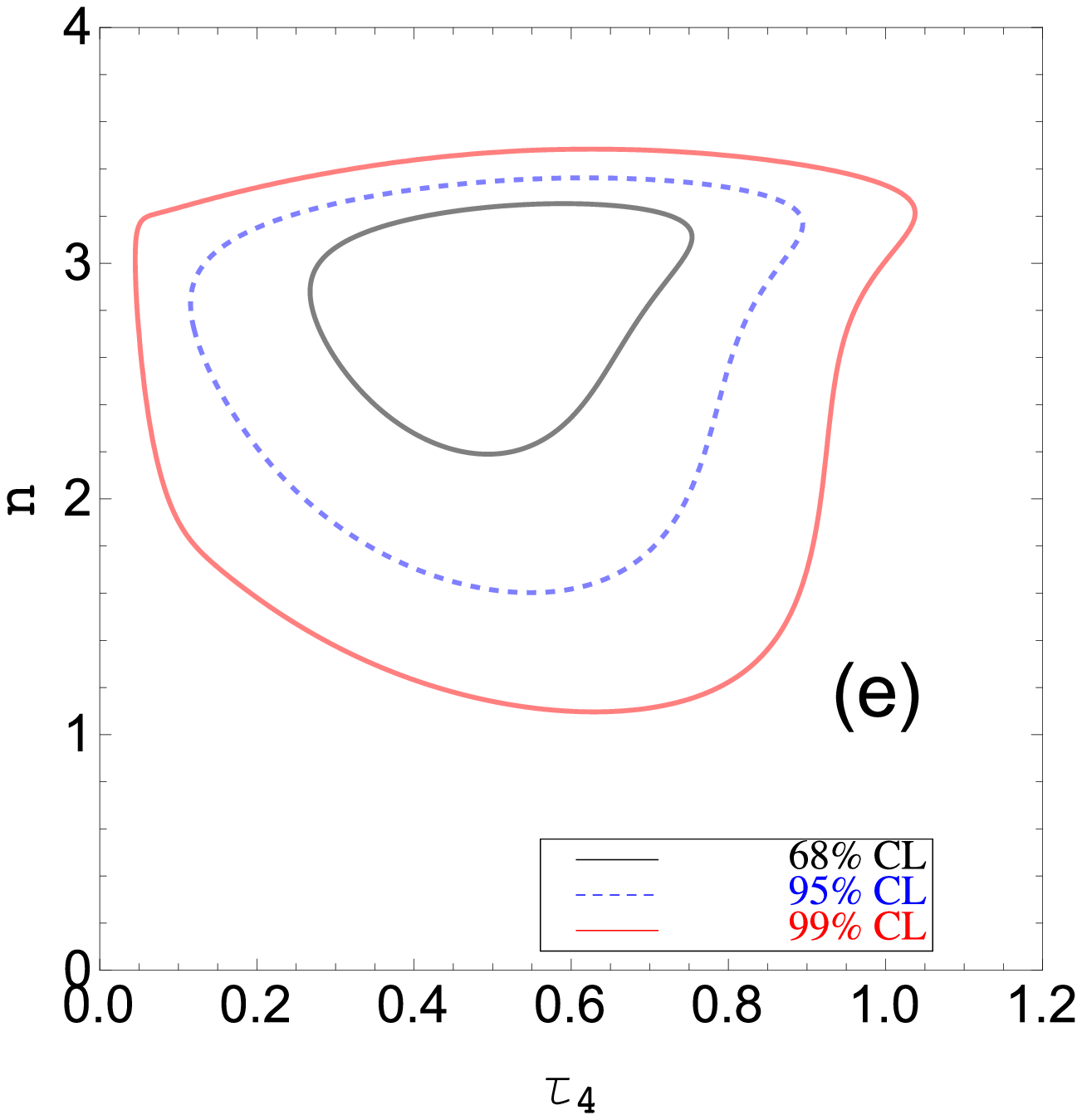}
\includegraphics[width=2.0in]{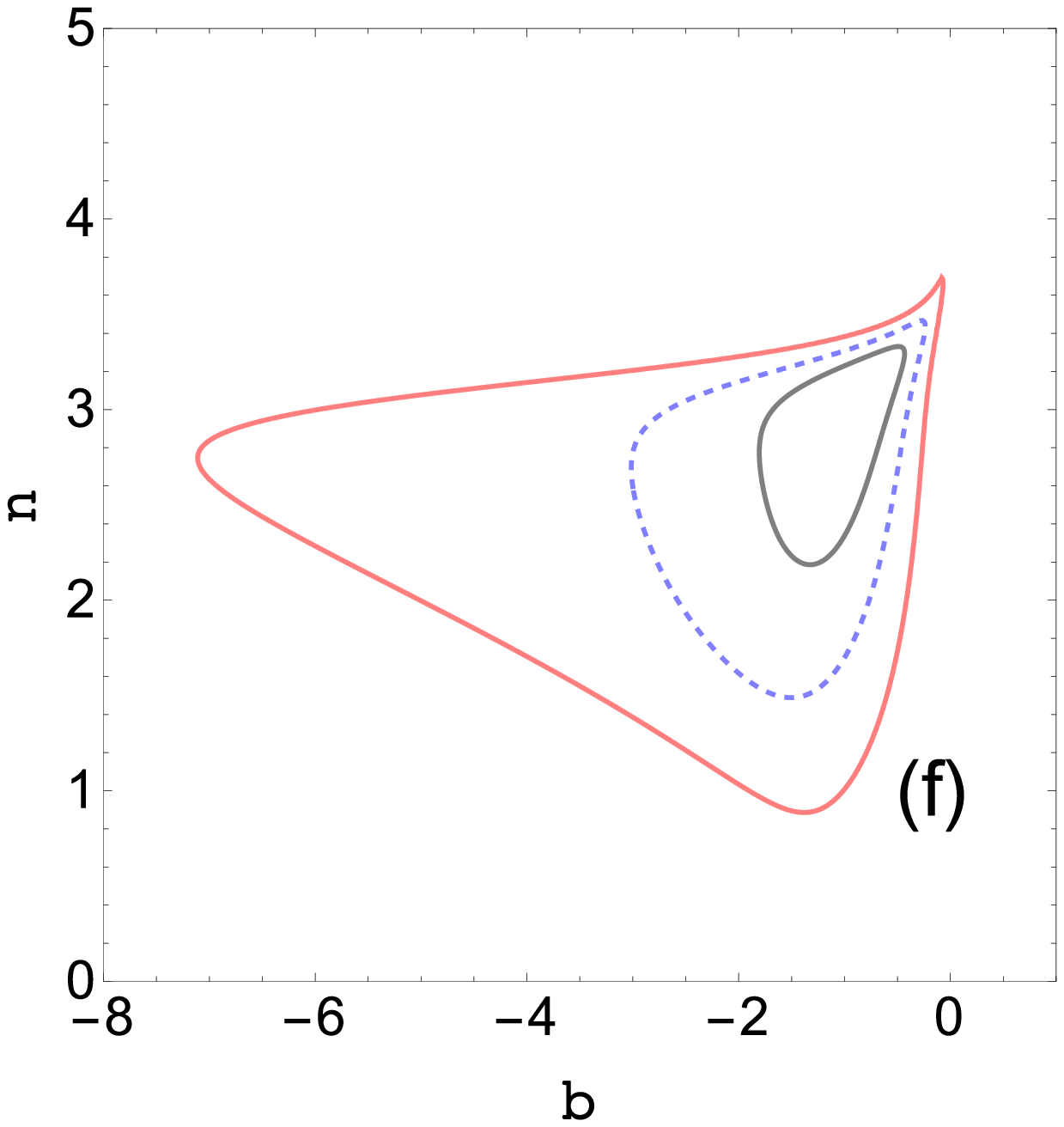}}
\caption{The fitting results for the varying power law model ($\tau,b,n$). {\it Panels (a), (b), (c)}: the marginalized distribution function of $\tau$, $b$, and $n$ respectively. For these three panels, the black dashed, red dashed, green dashed, and orange dashed lines are for {\tt SEVEM}, {\tt SMICA}, {\tt NILC} and {\tt COMMANDER} maps respectively, and the blue solid line is for the average $\chi^{2}$ of the previous four data sets (legend is shown in panel (b)). {\it Panels (d), (e), (f)}: the two-dimensional contours of joint constraints of ($\tau,b$), ($\tau,n$), and ($b,n$) respectively from the averaged $\chi^{2}$ function. The black, blue dashed and red lines are $68\%$, $95\%$ and $99\%$ C.L. respectively (legend is shown in panel (e)).}
\label{fig:fit_3p}
\end{figure*}

\begin{figure*}
\centerline{
\includegraphics[width=3.0in]{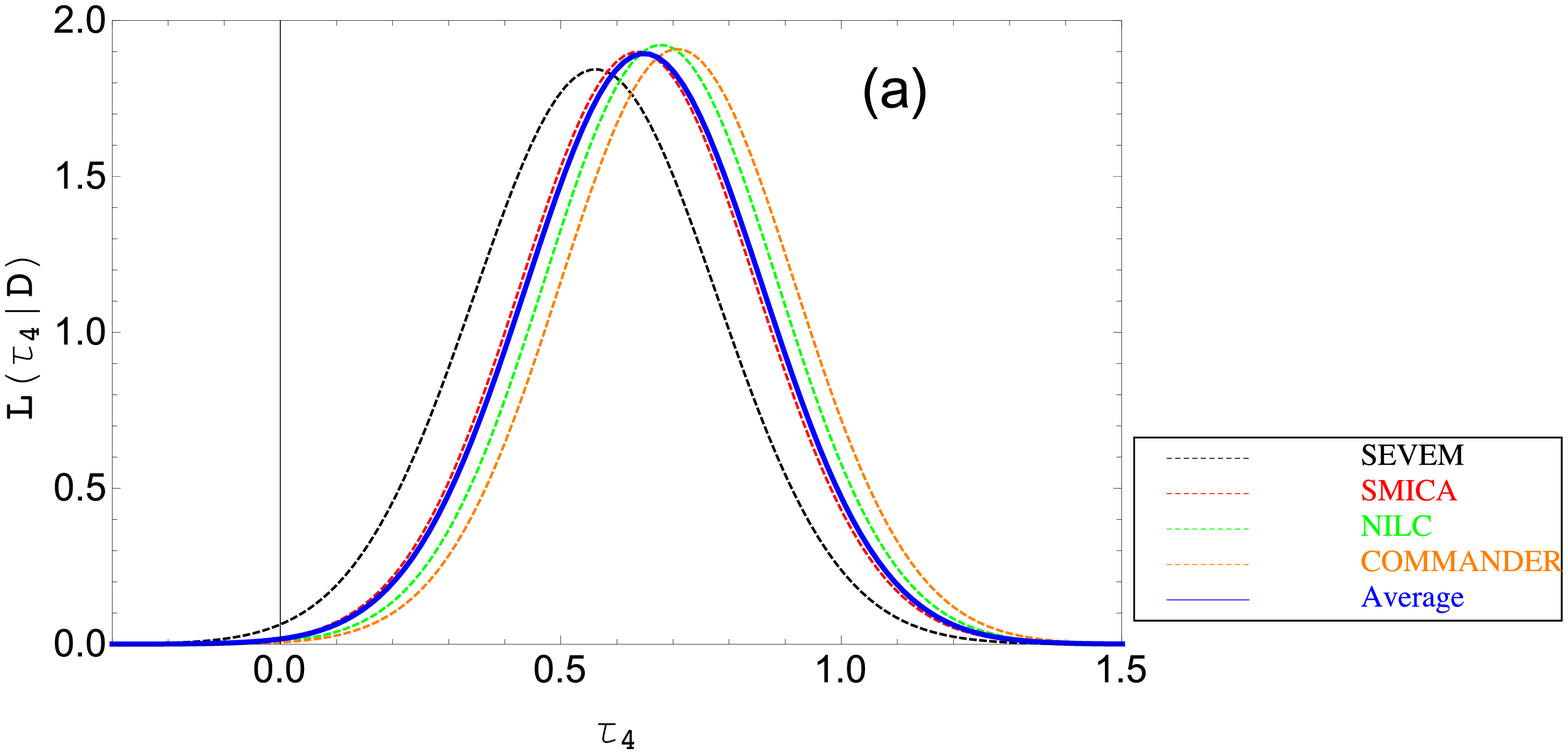}
\includegraphics[width=2.1in]{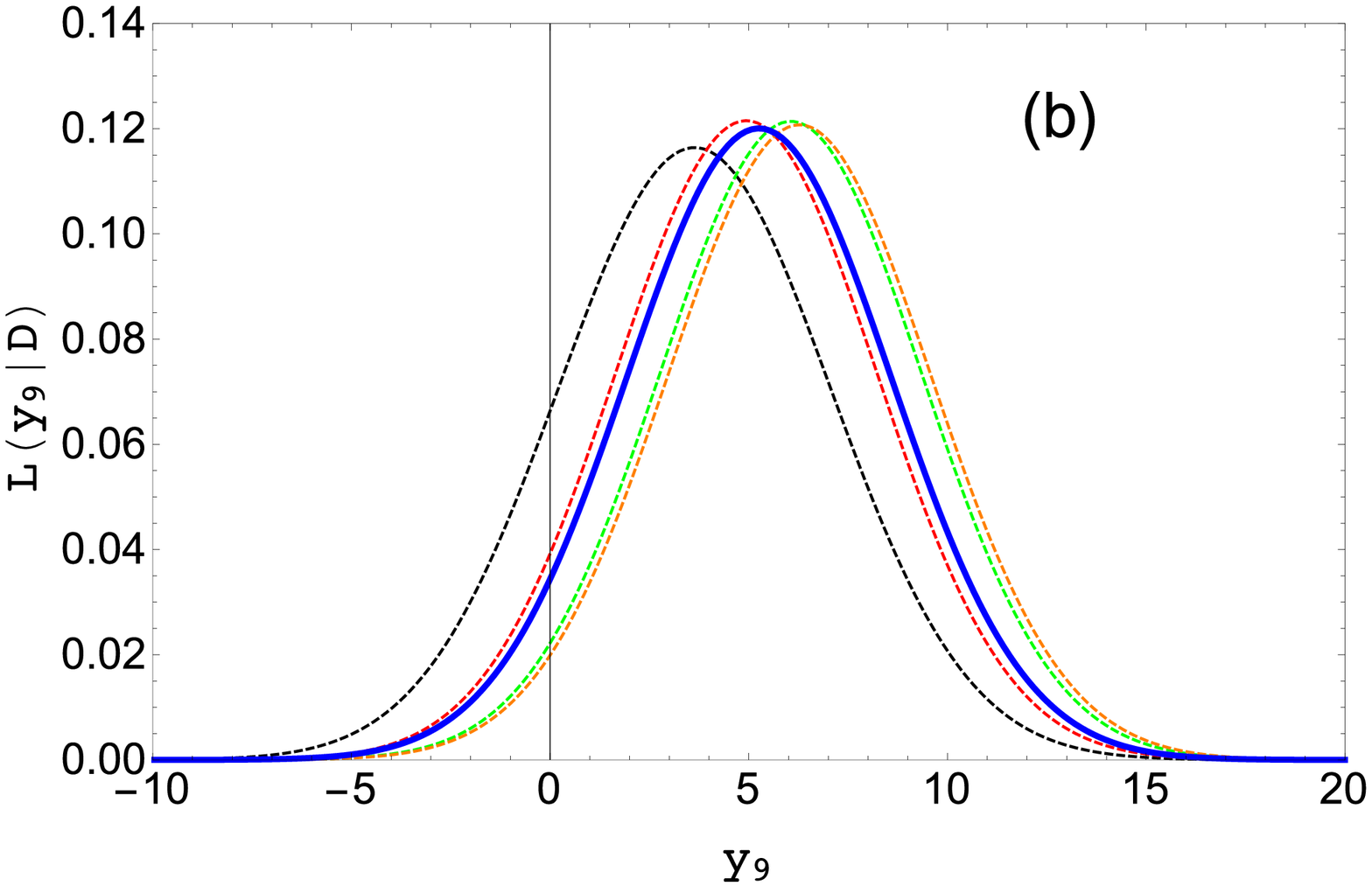}
\includegraphics[width=1.6in]{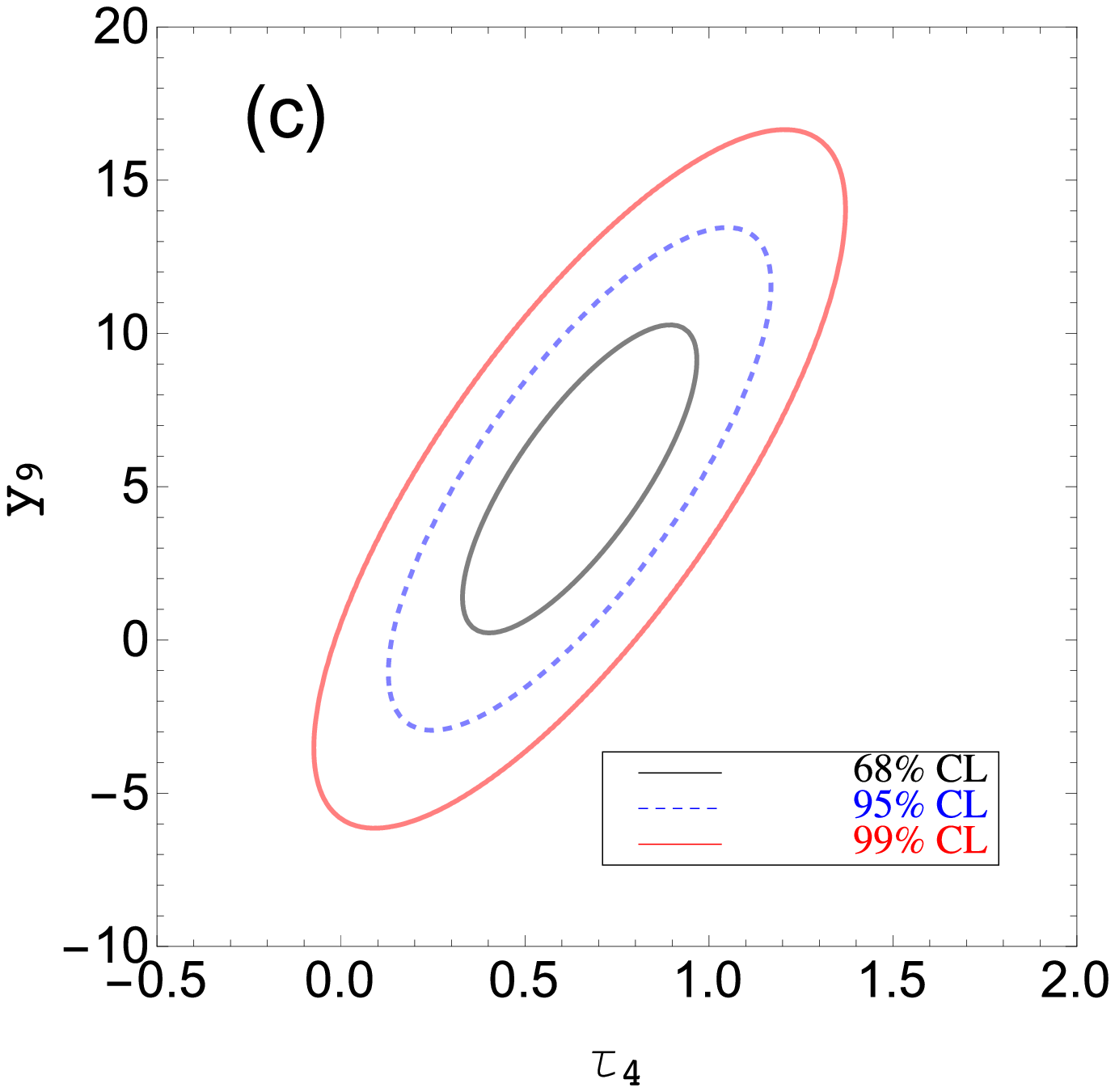}}
\caption{The fitting results for the unbiased model with calibration parameter, i.e. ($\tau,y_{0}$) model. {\it Panel (a)}: the marginalized distribution function of $\tau_{4}$; {\it Panel (b)}: the marginalized distribution function of $y_{9}$. In panels (a) and (b), the black dashed, red dashed, green dashed, and orange dashed lines are for {\tt SEVEM}, {\tt SMICA}, {\tt NILC} and {\tt COMMANDER} maps respectively. The blue solid line is for the average $\chi^{2}$ of the previous four data sets. {\it Panel (c)}: the joint constraint on $(\tau_{4},y_{9})$ from the averaged $\chi^{2}$ function.} \label{fig:scaling-2p}
\end{figure*}

\begin{figure*}
\centerline{
\includegraphics[width=2.2in]{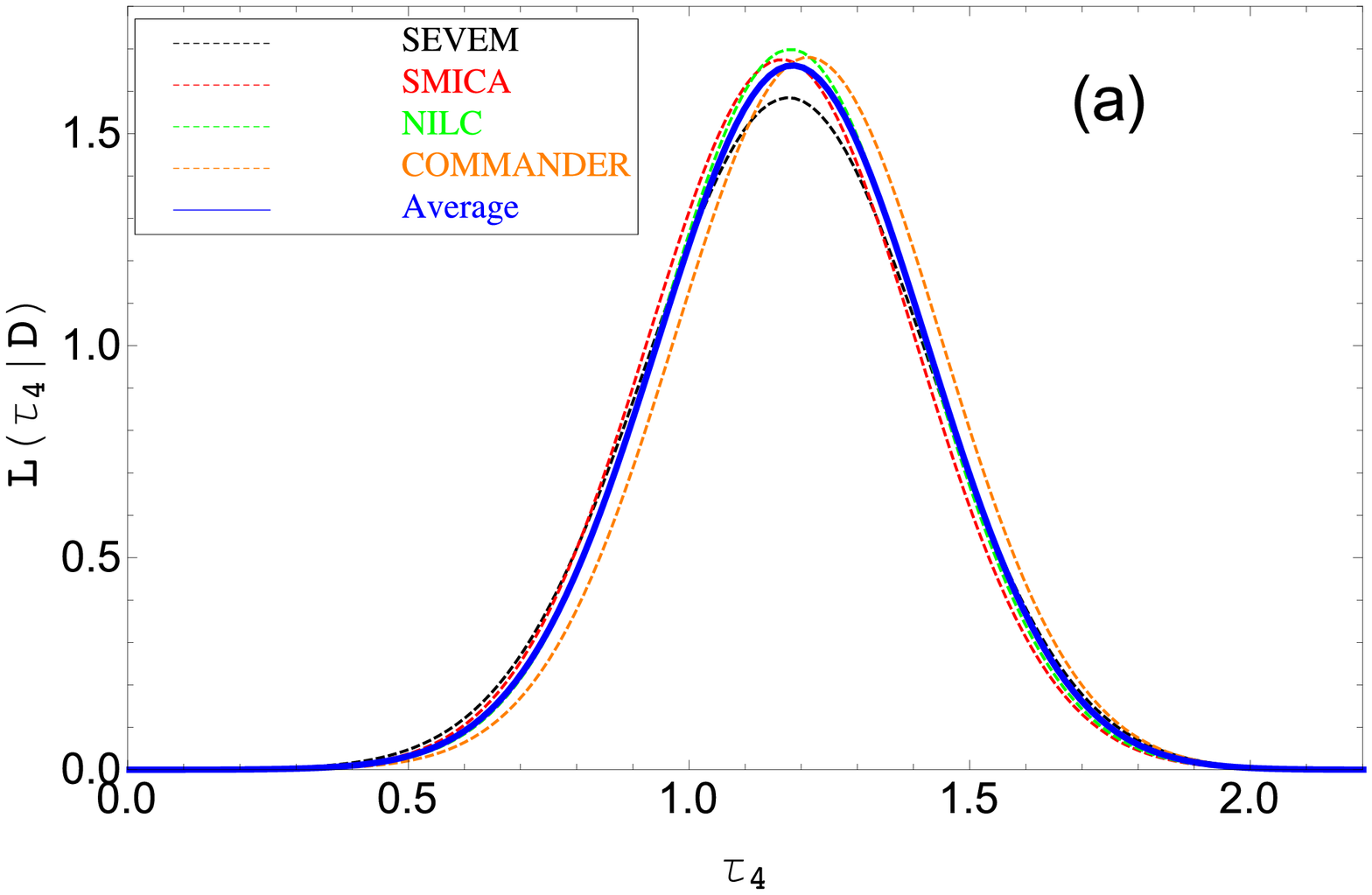}
\includegraphics[width=2.2in]{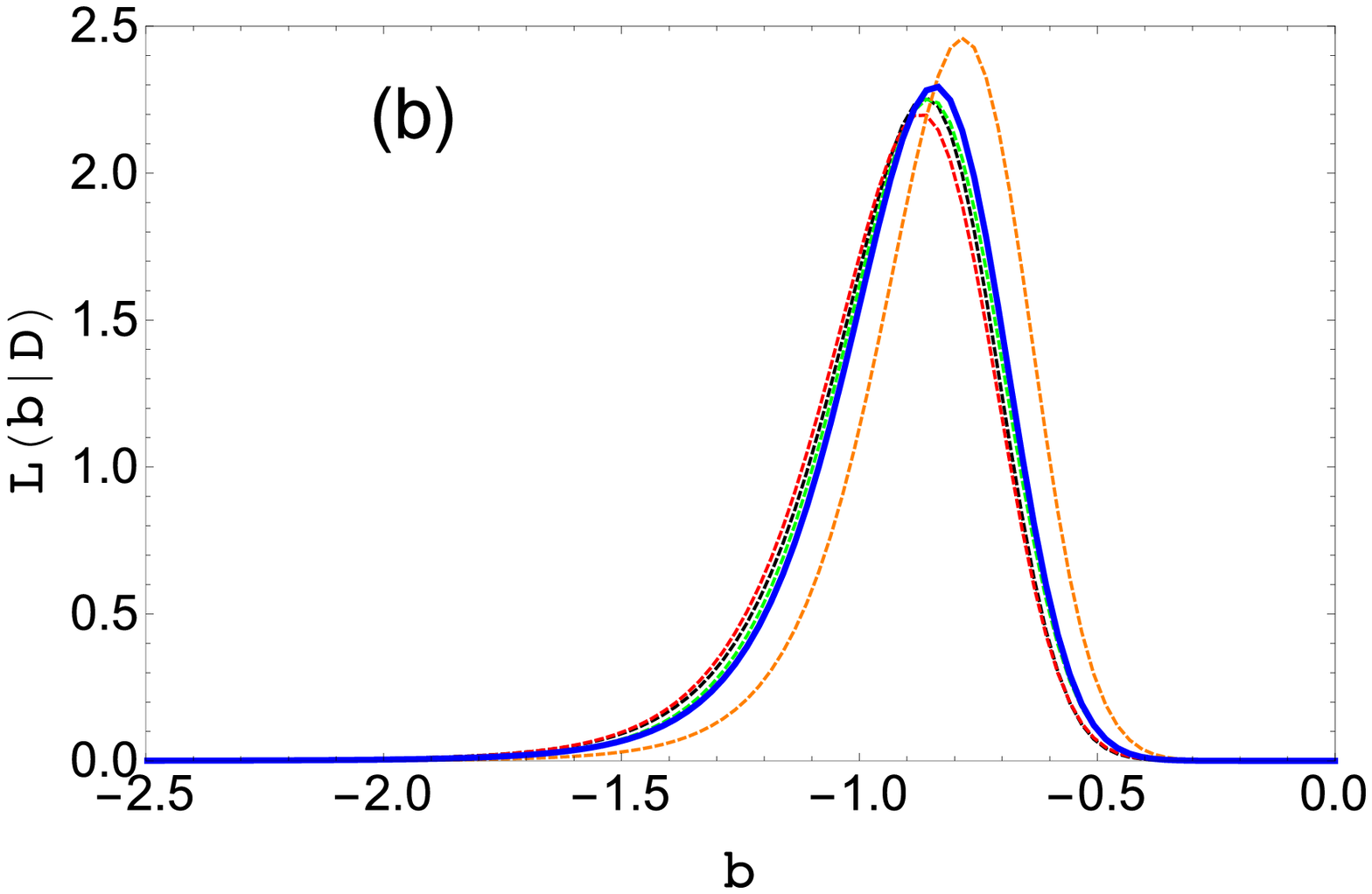}
\includegraphics[width=2.2in]{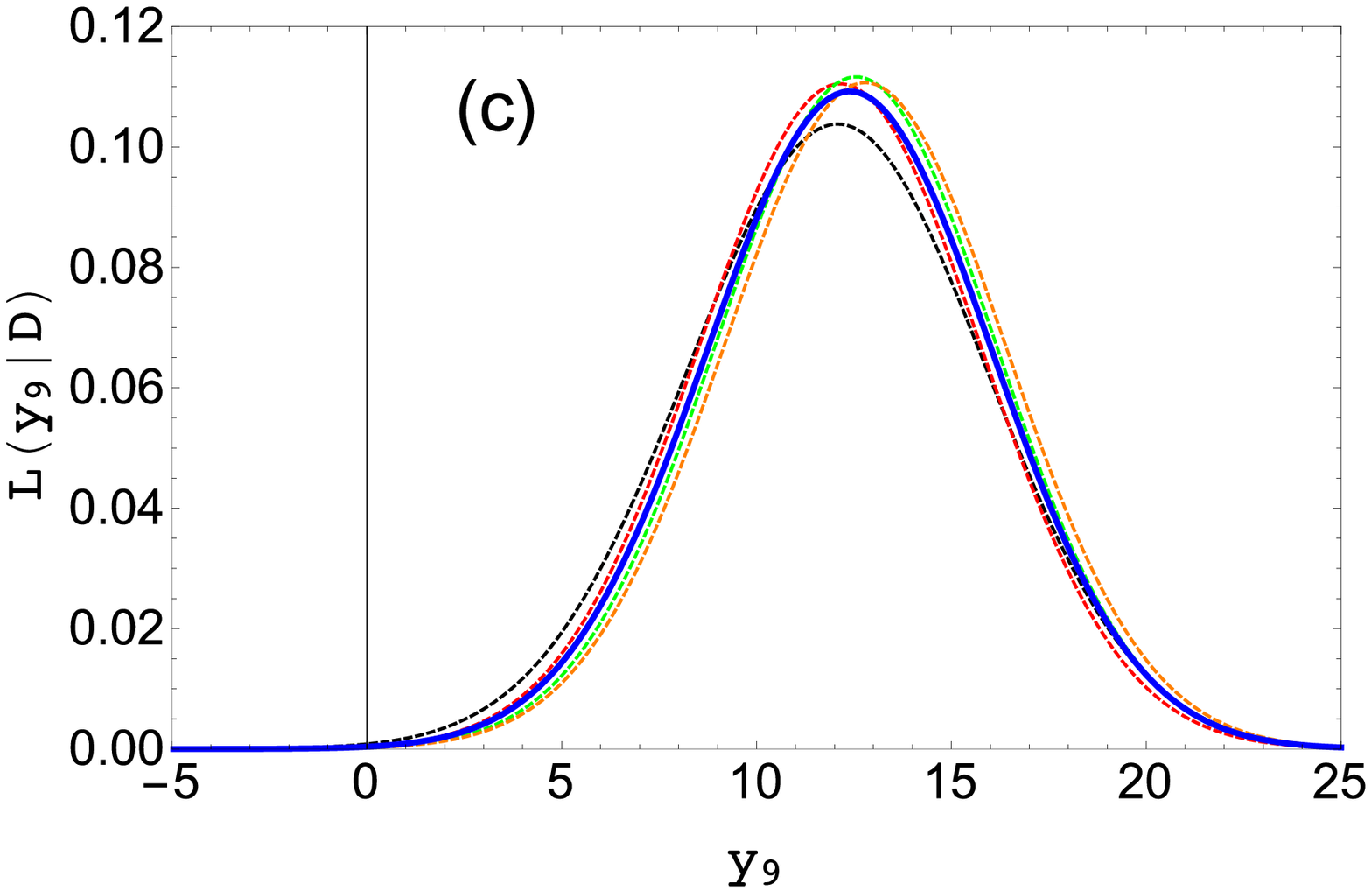}}
\centerline{
\includegraphics[width=2.2in]{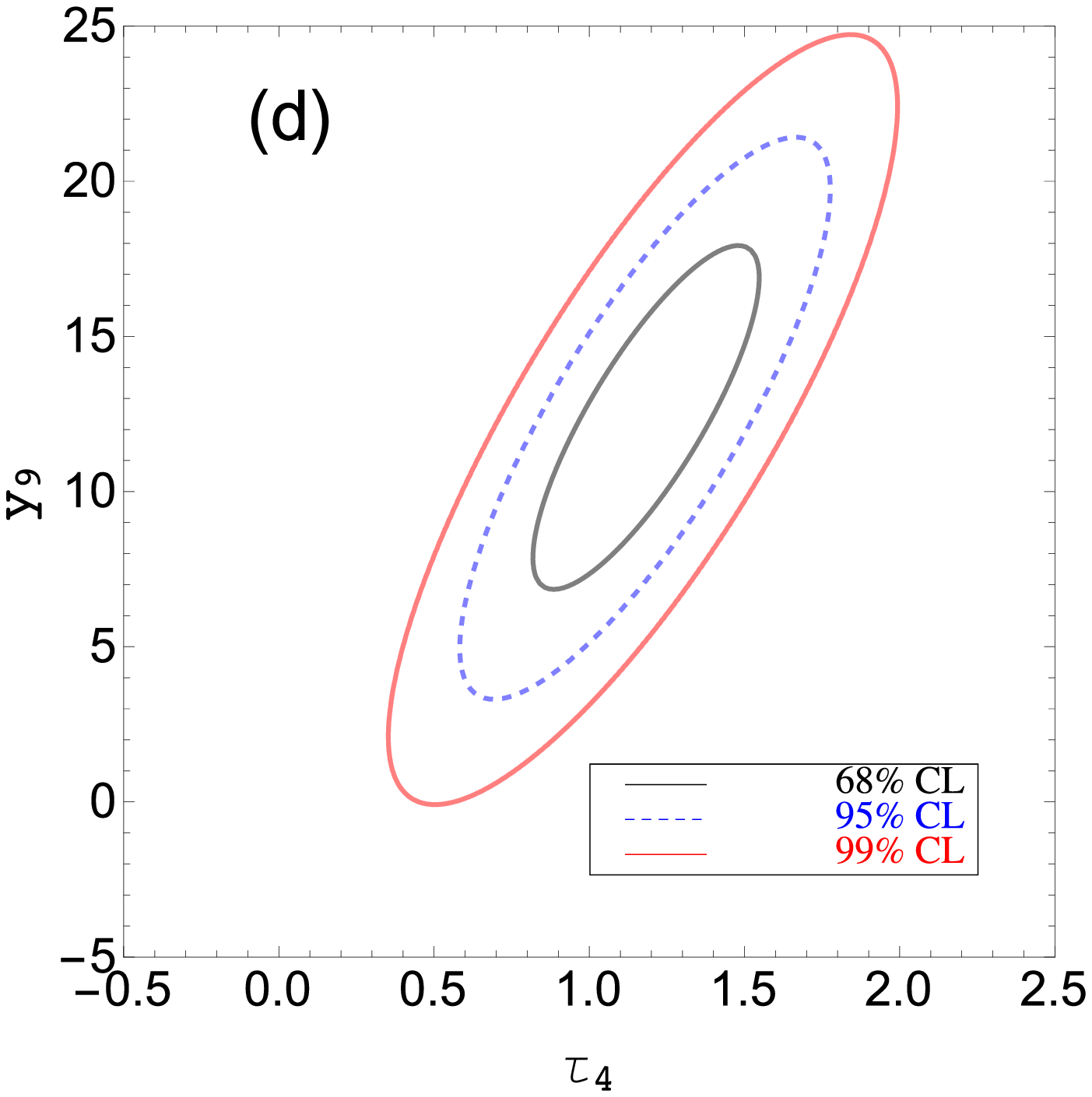}
\includegraphics[width=2.2in]{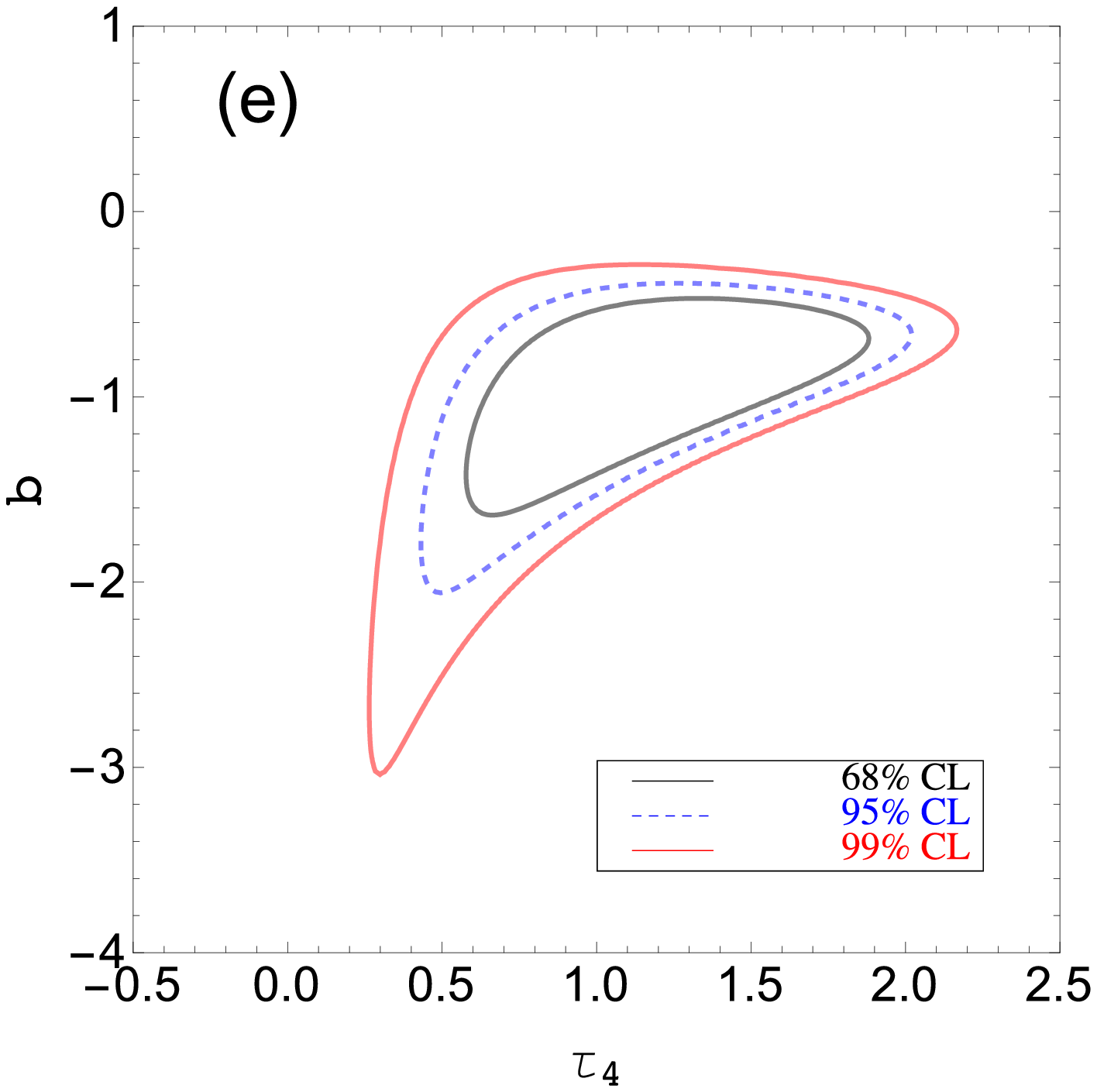}
\includegraphics[width=2.2in]{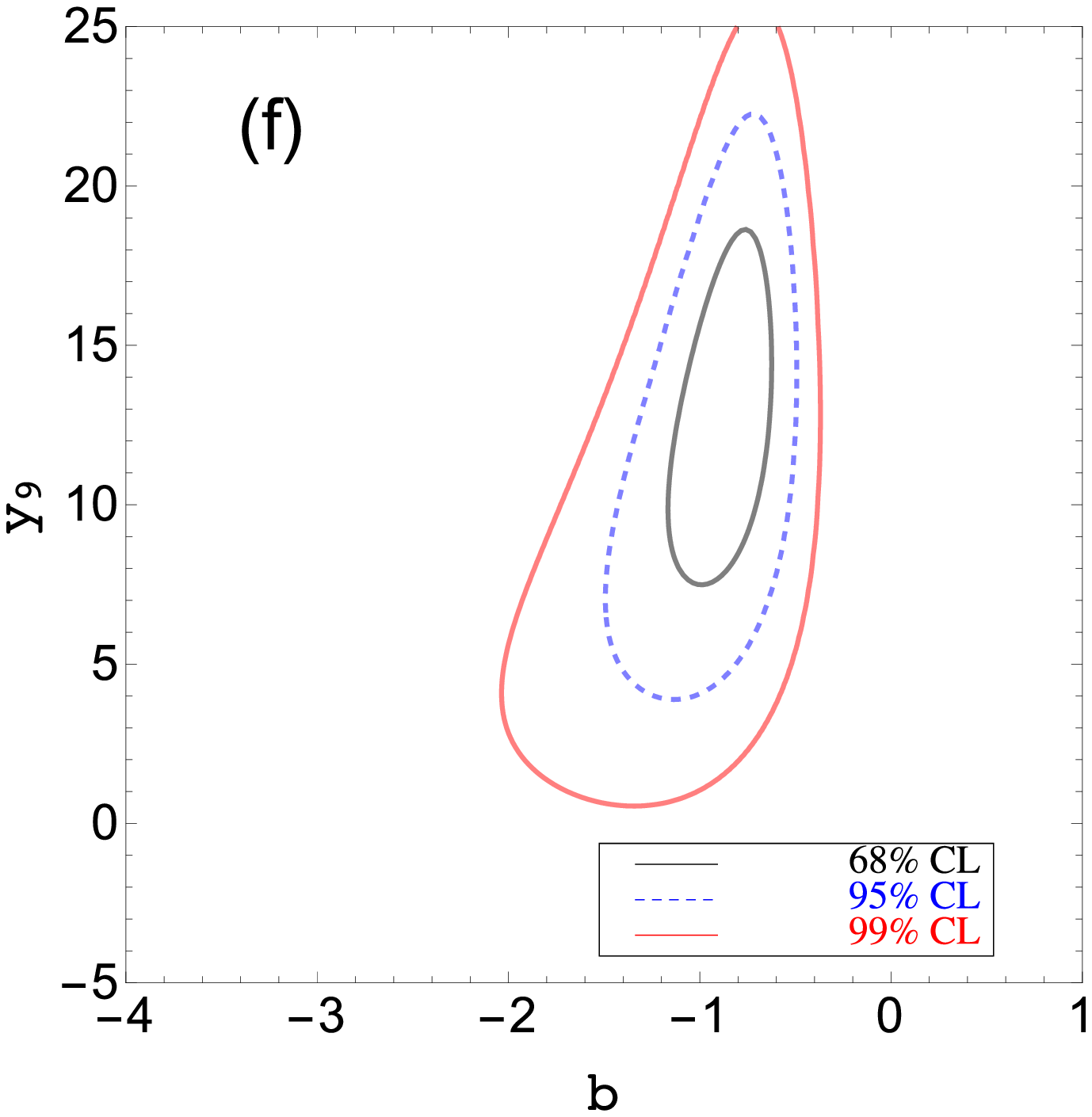}}
\caption{The fitting results for the quadratic power-law model with calibration parameter, i.e. ($\tau,b,y_{0}$) model. {\it Panels (a), (b), (c)}: the marginalized distribution function of $\tau_{4}$, $b$ and $y_{9}$ respectively. In panels (a), (b) and (c), the black dashed, red dashed, green dashed, and orange dashed lines are for {\tt SEVEM}, {\tt SMICA}, {\tt NILC} and {\tt COMMANDER} maps respectively. The blue solid line is for the average $\chi^{2}$ of the previous four data sets. {\it Panels (d), (e), (f)}: the joint constraints on ($\tau_{4}$,$y_{9}$), ($\tau_{4}$,$b$) and ($b$,$y_{9}$) from the averaged $\chi^{2}$ function.} \label{fig:scaling-3p}
\end{figure*}

\begin{figure*}
\centerline{
\includegraphics[width=2.6in]{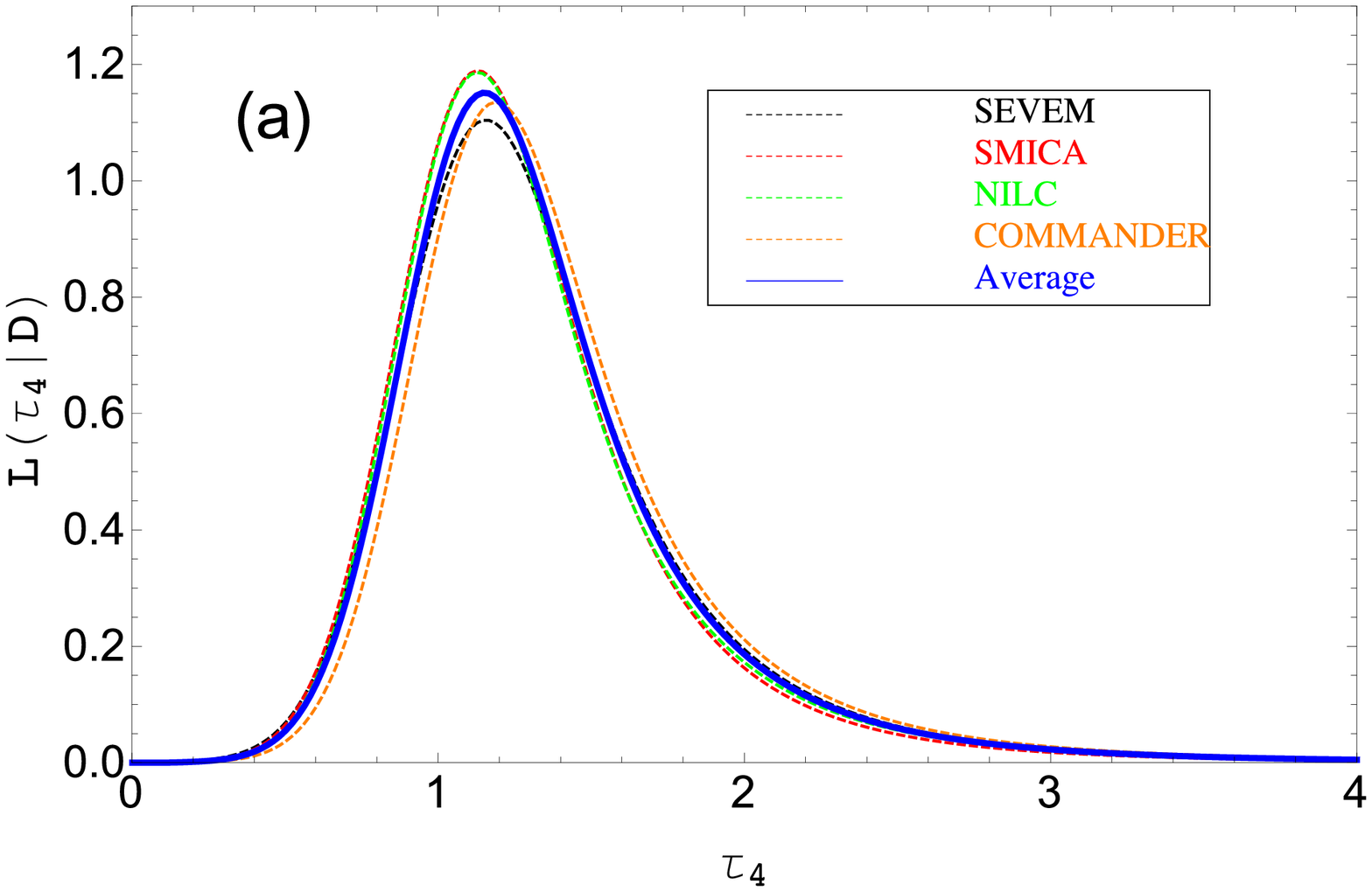}
\includegraphics[width=2.5in]{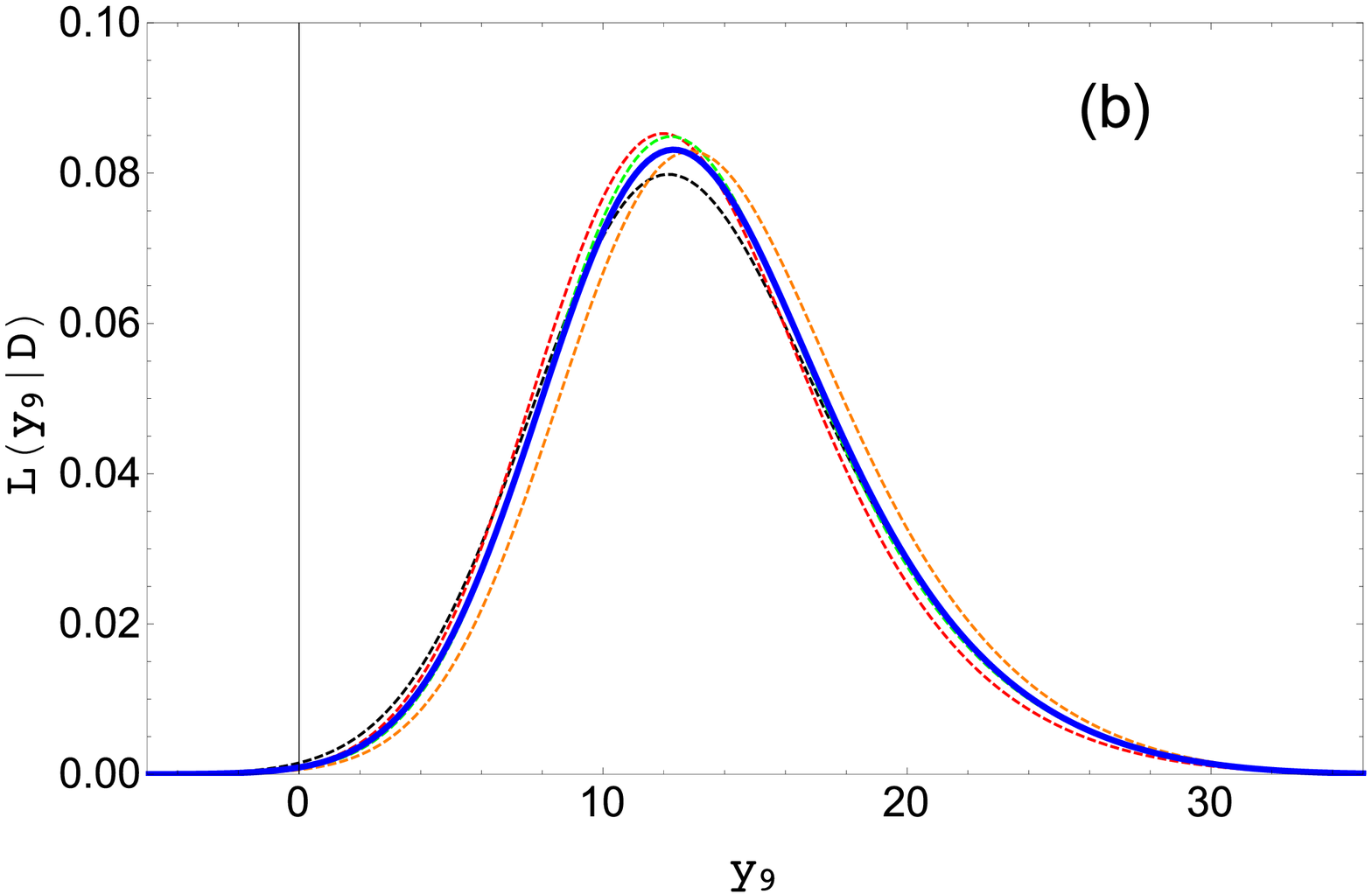}}
\centerline{
\includegraphics[width=2.5in]{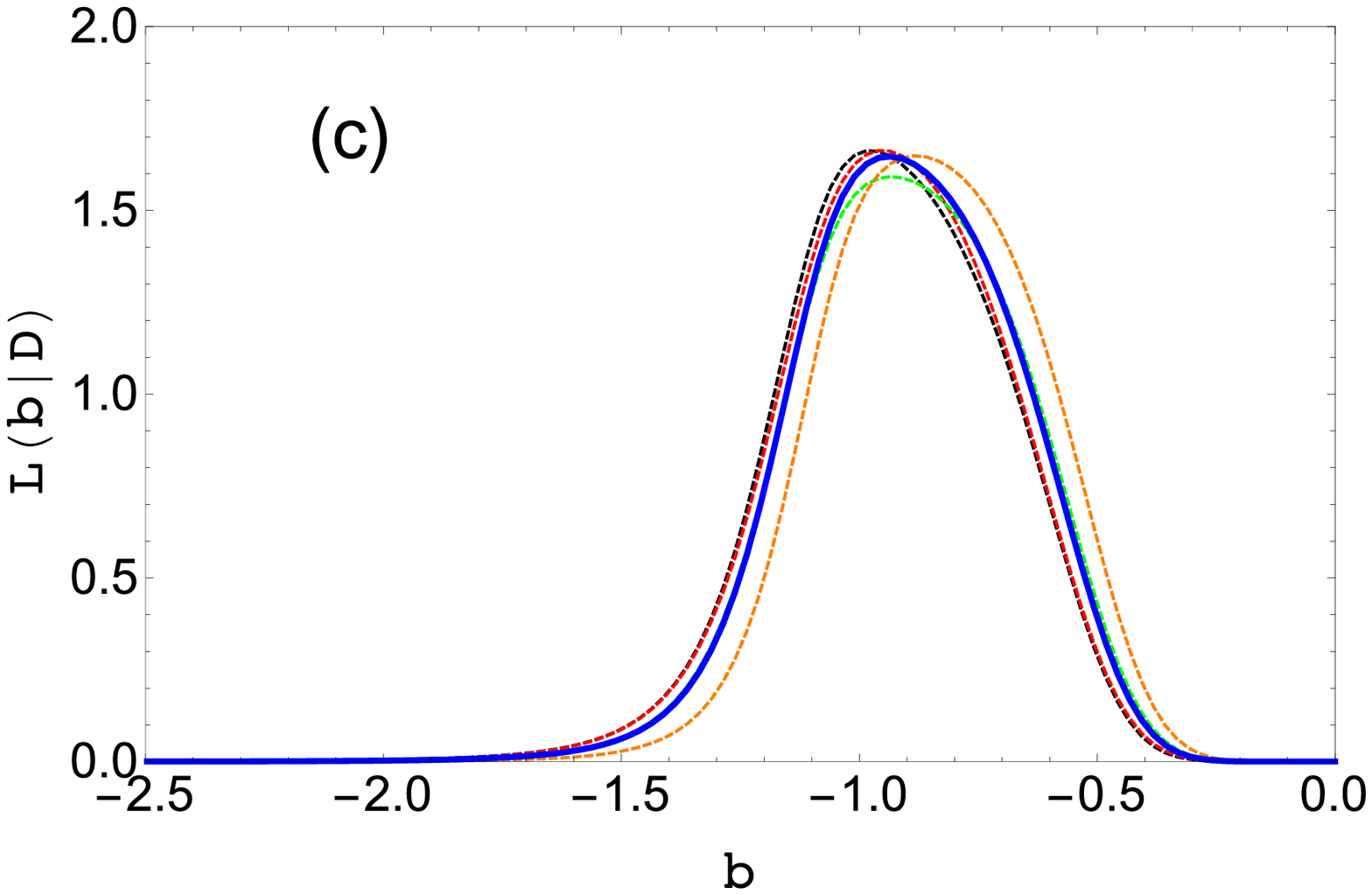}
\includegraphics[width=2.5in]{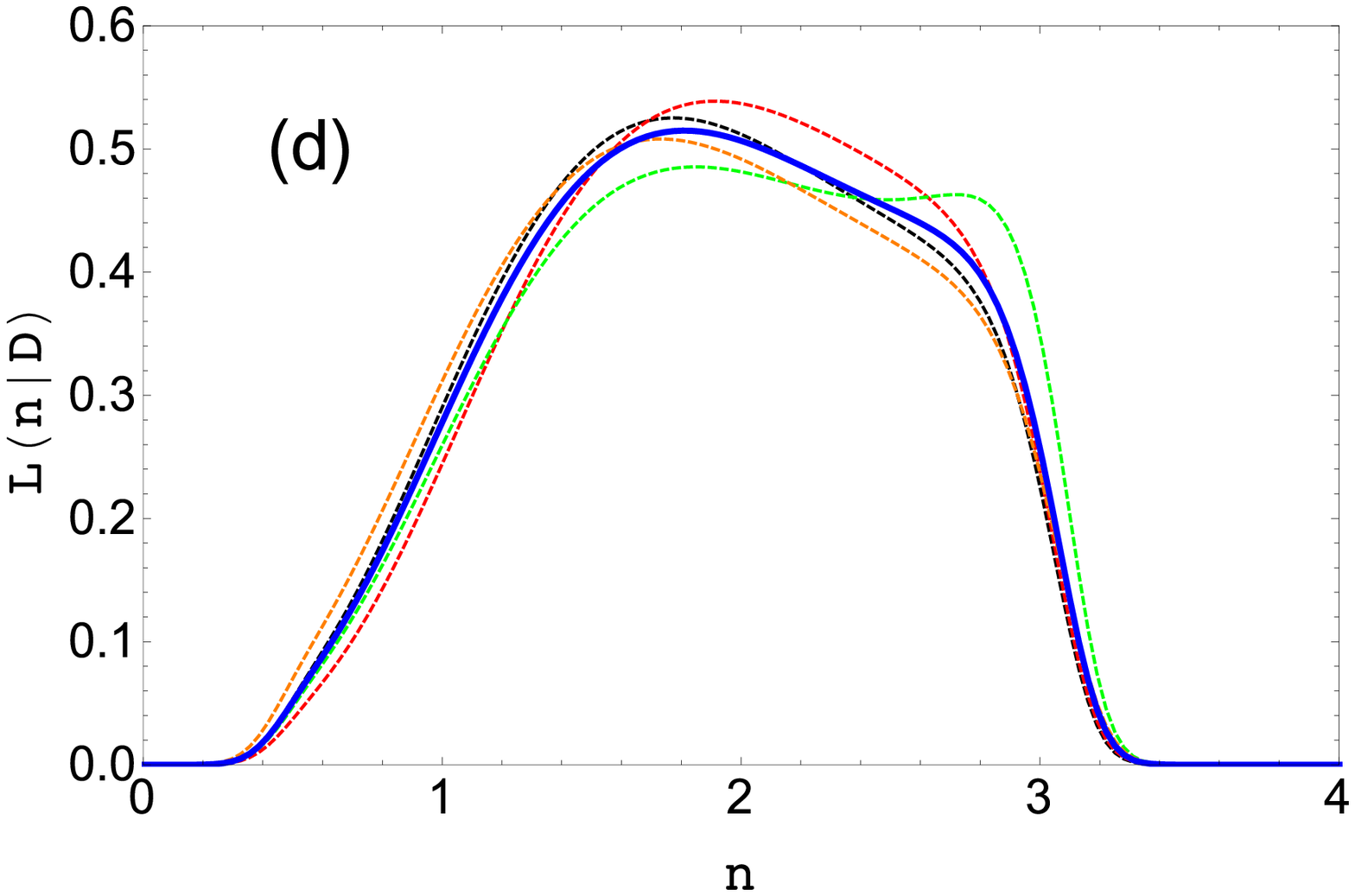}}
\caption{The fitting results for the varying power-law model with calibration parameter, i.e. ($\tau,b,n,y_{0}$) model. Legend is the same as Fig.~\ref{fig:scaling-3p}.} \label{fig:scaling-4p}
\end{figure*}

\begin{figure*}
\centerline{
\includegraphics[width=3.0in]{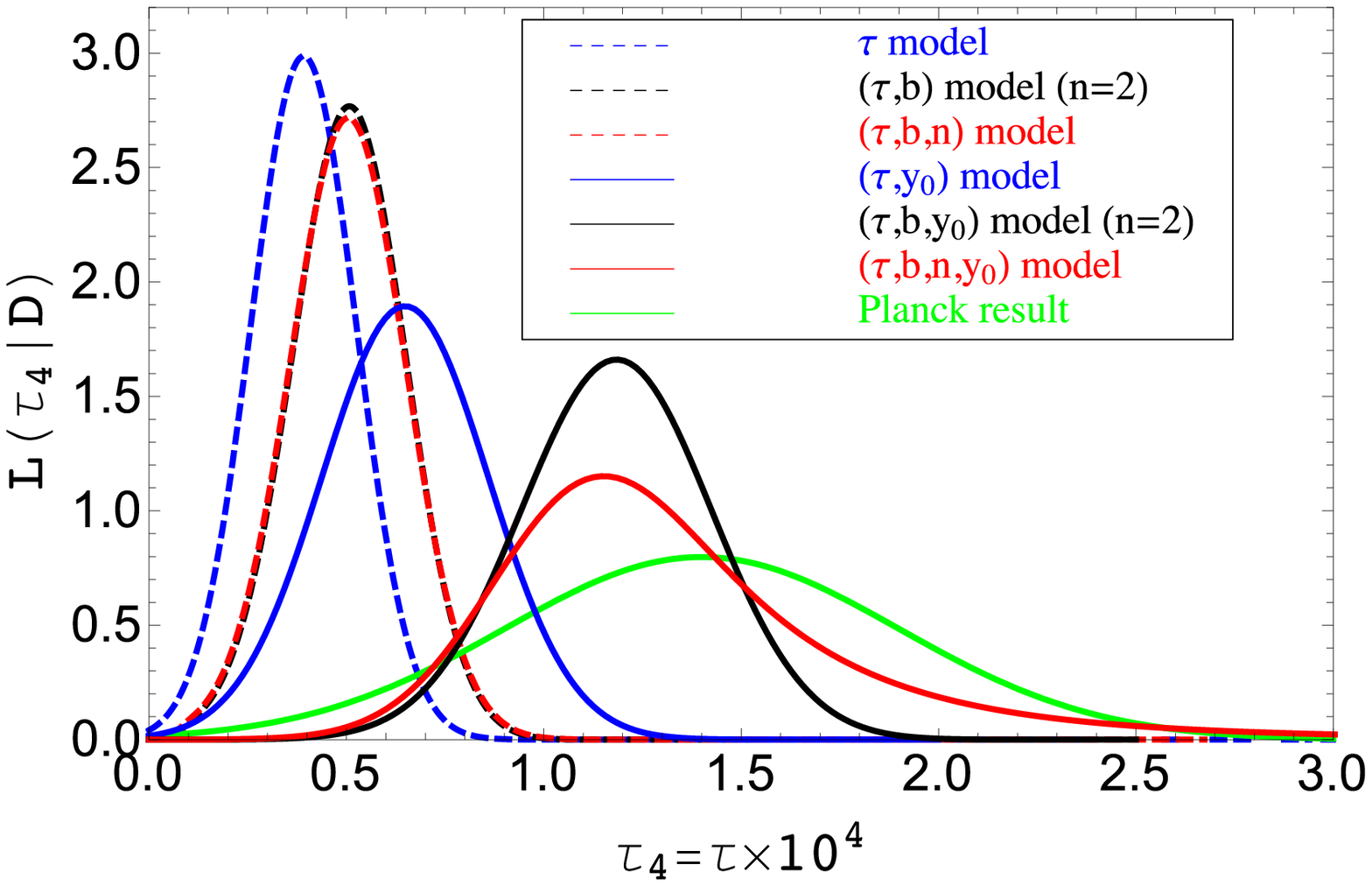}
\includegraphics[width=3.0in]{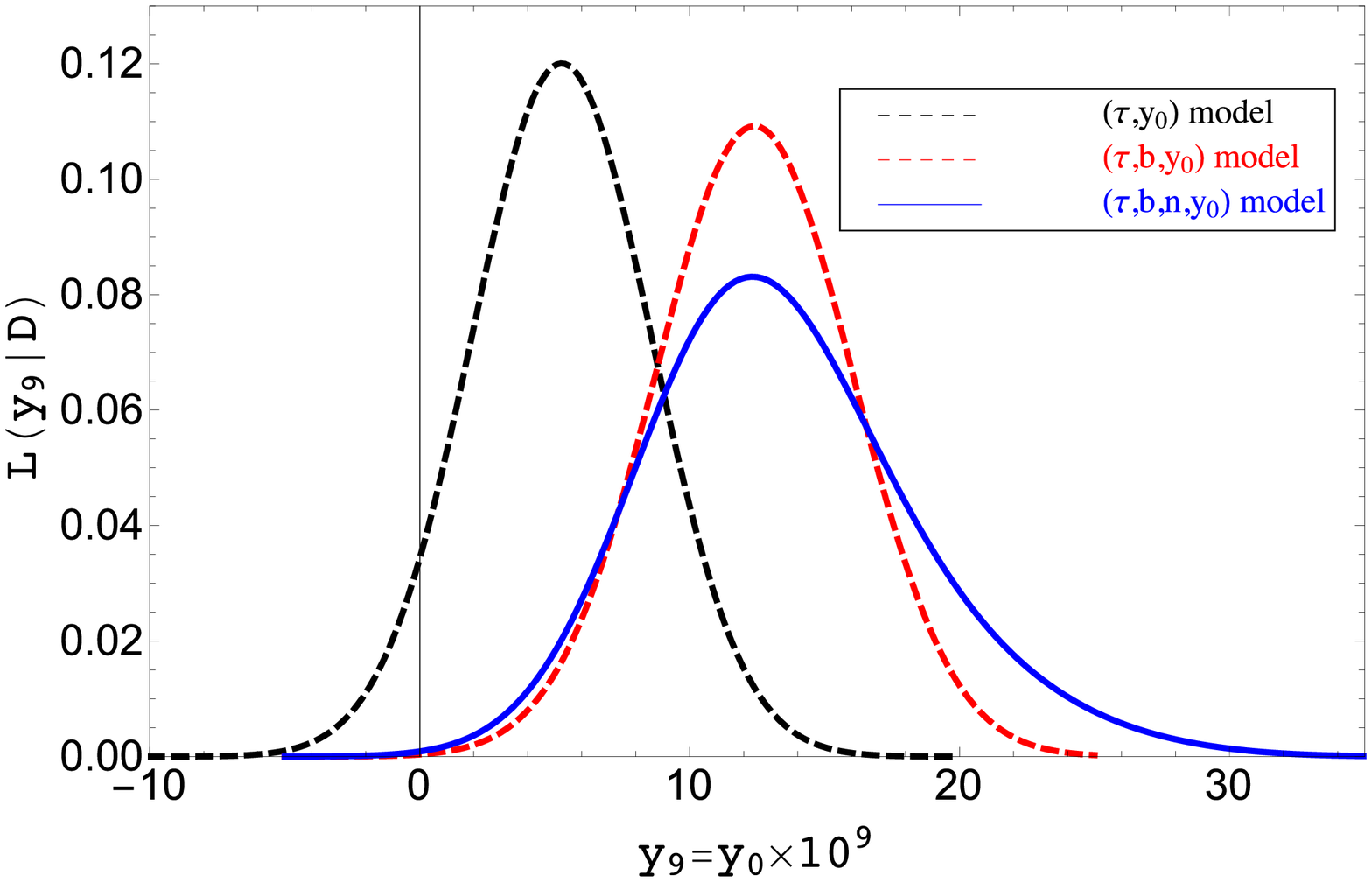}}
\caption{{\it Left}-- Comparison between distribution function of $\tau$ for the six models studied in this paper and the {\it Planck} result~\citep{Planck16-unbound}. The {\it Planck} result is obtained by fitting the $\langle \Delta T v \rangle$ data of {\tt SEVEM} map with the template from N-body simulation. The results between the $(\tau,b,y_{0})$ model and $(\tau,b,n,y_{0})$ model and {\it Planck} are consistent with each other within $1\sigma$ C.L. {\it Right}-- Comparison of the fitting results between different models.} \label{fig:comp}
\end{figure*}

\subsection{Cross-correlation}

We now calculate the correlation between kSZ temperature distortion and the linear velocity field. For any pair of galaxies at spatial point $\mathbf{r}_{1}$ and $\mathbf{r}_{2}$ (comoving coordinates), their correlation is
\begin{eqnarray}
\left\langle \left(\frac{\Delta T}{T} \right) \left(\frac{\mathbf{v}\cdot \hat{\mathbf{n}}}{\sigma_{v}} \right) \right\rangle =-\left(\frac{\tau}{c \sigma_{v}} \right) \langle (\mathbf{v}_{1}\cdot \mathbf{\hat{n}}_{1})(\mathbf{v}_{2}\cdot \mathbf{\hat{n}}_{2}). \rangle \label{eq:deltaT-v}
\end{eqnarray}

The angular average of the right-hand-side of Eq.~(\ref{eq:deltaT-v}) should be negative, because has a negative sign in front of sample average of the quantity $\langle (\mathbf{v}_{1}\cdot \hat{\mathbf{n}}_{1}) (\mathbf{v}_{2}\cdot \hat{\mathbf{n}}_{2}) \rangle$. The ensemble average of the line-of-sight velocity field is
\begin{eqnarray}
&& \langle (\mathbf{v}_{1}\cdot \mathbf{\hat{n}}_{1})(\mathbf{v}_{2}\cdot \mathbf{\hat{n}}_{2}) \rangle \nonumber \\
&& =
\frac{1}{(2\pi)^{3}} \int \der^{3} \mathbf{k} P_{v}(k)(\mathbf{\hat{k}}\cdot \mathbf{\hat{n}}_{1})(\mathbf{\hat{k}}\cdot \mathbf{\hat{n}}_{2})
\me^{i \mathbf{k} \cdot (\mathbf{r}_{1}-\mathbf{r}_{2})} \nonumber \\
&& = \frac{1}{2 \pi^{2}} \int \der k k^{2}P_{vv}(k) F(k;\mathbf{r}_{1},\mathbf{r}_{2}) \nonumber \\
&& = \frac{a^{2}H^{2}f(z_1)f(z_2)}{2\pi^{2}} \int \der k b_{v}(k) P_{\rm m}(k)F(k;\mathbf{r}_{1},\mathbf{r}_{2}), \label{eq:vrvr-1}
\end{eqnarray}
where $P_{\rm m}(k)$ is the matter power spectrum at redshift zero, and $f(z)=\der \ln D/\der \ln a$ is the growth factor. In Eq.~(\ref{eq:vrvr-1}), we defined the peculiar velocity bias function $b_{v}(k)$ as
\begin{eqnarray}
P_{vv}(k)=b_{v}(k)\left(\frac{a^{2}H^{2}f^{2}(z)}{k^{2}} \right)P_{\rm m}(k), \label{eq:bv-def}
\end{eqnarray}
and the angular integration to be
\begin{eqnarray}
F(k;\mathbf{r}_{1},\mathbf{r}_{2})=\frac{1}{4 \pi} \int \der^{2}\hat{\mathbf{k}}\,(\hat{\mathbf{k}}\cdot \hat{\mathbf{n}}_{1})(\hat{\mathbf{k}}\cdot \hat{\mathbf{n}}_{2}){\rm e}^{i \mathbf{k}\cdot (\mathbf{r}_{1}-\mathbf{r}_{2})}.
\end{eqnarray}
Note that if the velocity completely traces the underlying matter distribution, $b_{v}(k) \equiv 1$, Eq.~(\ref{eq:bv-def}) reduces to the prediction of linear perturbation theory.

The function $F(k;\mathbf{r}_{1},\mathbf{r}_{2})$ depends only on the magnitude $r_{1}$, $r_{2}$ and the cosine angle between the two vectors. The appendix in~\citet{Ma11} showed that it can be solved analytically as
\begin{eqnarray}
F(k; \mathbf{r}_{1},\mathbf{r}_{2}) &=& \frac{1}{3}\cos \alpha (j_{0}(kr)-2j_{2}(kr)) \nonumber \\
&+&  \frac{j_{2}(kr)}{r^{2}}r_{1}r_{2}\sin^{2}\alpha, \label{eq:Fr_k}
\end{eqnarray}
where $\cos \alpha$ is the cosine angle between the two vectors, and $j_0$ and $j_2$ are spherical Bessel functions of order zero and two, and $r=\left(r^{2}_{1}+r^{2}_{2}-2r_{1}r_{2}\cos\alpha \right)^{1/2}$ is the separation distance between the two galaxies. Therefore, one can see that the average of the line-of-sight peculiar velocity field (Eq.~(\ref{eq:vrvr-1})) depends on the individual distances of the two galaxies and their cosine angle, and it preserves statistical isotropy but {\it not} statistical homogeneity. This is because, although the three-dimensional velocity field ($\mathbf{v}$) itself is both statistical homogeneous and isotropic (the irrotational velocity field is a gradient of a statistically symmetric scalar field), once a specific point (the observer's location) is chosen, not all pairs of galaxies of a given vectorial separation have the same relation to this origin. In this sense, we need to calculate $\langle (\Delta T/T)(\mathbf{v}\cdot \hat{\mathbf{n}}/\sigma_{v}) \rangle$ for the particular catalogue we use.

As one can see in Fig.~\ref{fig:deltaTv}, the data are plotted as the average signal at each different distance-separation bin. Therefore we need to calculate the average correlation function for different distance bins for the catalogue, i.e. we need to calculate
\begin{eqnarray}
y(r) & \equiv &  \overline{\left\langle \left(\frac{\Delta T}{T} \right) \left(\frac{\mathbf{v}\cdot \hat{\mathbf{n}}}{\sigma_{v}} \right) \right\rangle} \nonumber \\
& = & \frac{1}{N_{\rm pair}}\sum^{\rm all\,pairs}_{i,j} \left\langle \left(\frac{\Delta T(\hat{\mathbf{n}_{i}})}{T} \right) \left(\frac{\mathbf{v}\cdot \hat{\mathbf{n}_{j}}}{\sigma_{v}} \right) \right\rangle \nonumber \\
& = & -\left(\frac{\tau}{c\sigma_{v}} \right)\frac{a^{2}H^{2}f^{2}(z)}{2\pi^{2}}\int \der k\,b_{v}(k)P_{\rm m}(k)F_{r}(k), \label{eq:deltaTv-3}
\end{eqnarray}
where
\begin{eqnarray}
F_{r}(k)=\frac{1}{N_{\rm pair}}\sum^{\rm all\, pairs}_{ij}F(k; \mathbf{r}_{i},\mathbf{r}_{j}), \label{eq:Frk2}
\end{eqnarray}
for all pairs whose separation is in between $r-\Delta r$ and $r+\Delta r$.

We calculate the averaged angular function (Eq.~(\ref{eq:Frk2})) for all galaxy pairs at each of the separation distance bins $[r-\Delta r, r+\Delta r]$. We plot the resulting functions in Fig.~\ref{fig:Frk}. One can see that on large scales, as $k\rightarrow 0$ and $r\rightarrow 0$, the spherical Bessel function $j_{0}(kr)\rightarrow 1$, $j_{2}(kr) \rightarrow 0$ and $\cos \alpha \rightarrow 1$, so the function approaches $1/3$ (Eq.~(\ref{eq:Fr_k})). But on small scales $k \rightarrow 1\,h^{-1}$Mpc the averaged angular function highly oscillates and approaches zero at $k \gg 0.1\,h\,$Mpc.

The data plotted in Fig.~\ref{fig:deltaTv} are derived from {\it Planck} maps that have a uniform FWHM equal to $5\,$arcmin. So, in principle, we should convolve our correlation function with a Gaussian beam also on this angular scale. However, as one can see in the histogram (Fig.~\ref{fig:angle}), the distribution of averaged separation angle between pairs of galaxies peaks at $15\,$degree, much larger than the width of the {\it Planck} beam, so the Gaussian beam does not affect the calculation of the correlation function. For this reason, we can regard our source as an extended source which does not change very much after convolution with a small angular scale beam.

\subsection{Velocity bias models}
\label{sec:bias}
We vary the optical depth $\tau$ in the fitting, which controls the total amplitude of correlation function (Eq.~(\ref{eq:deltaTv-3})). Besides this we also vary the velocity bias function, and see whether it can improve the fitting of correlation function. 
In our model, the velocity bias relates the galaxy peculiar velocity to the underlying dark matter distribution, i.e. $v_{\rm g}=b_{v}\delta_{\rm m}$. The normal galaxy bias, which relates the density contrast of galaxies to dark matter ($\delta_{\rm g}=b_{\rm g}\delta_{\rm m}$), does not enter into this relation.

The velocity bias parameter has two limits: (1) As $k \rightarrow 0$, $b_v(k) \rightarrow 1$, because on large scales the velocity bias vanishes; (2) When $k$ becomes large, $|b_v(k)\gg 1|$. In this paper, we consider the following three models:
\begin{enumerate}

\item An unbiased model~\citep{Percival08,Zhang15,Zheng15a,Zheng15b}. In this model
\begin{eqnarray}
b_{v}(k) \equiv 1,
\end{eqnarray}
for all ranges of $k$. The only parameter to vary is $\tau$. Below, we name this model as the ``$\tau$ model'', or single-parameter model.

\item A quadratic power-law model~\citep{Baldauf15,Chan15}. In this model,
\begin{eqnarray}
b_{v}(k)=1+b\left(\frac{k}{k_{0}} \right)^{2}, \label{eq:bv-2}
\end{eqnarray}
where $b$ is varied. $k_{0}$ is the pivot scale fixed to be $0.1\,h\,{\rm Mpc}^{-1}$. Below, we name this model as the ``($\tau, b, n=2$) model''. This quadratic power-law model is expected in the standard peak background split theory~\citep{Desjacques08,Desjacques10,Baldauf15,Chan15}.

\item A varying power-law model~\citep{Percival08,Zhang15,Zheng15a,Zheng15b}. In this model,
\begin{eqnarray}
b_{v}(k)=1+b\left(\frac{k}{k_{0}} \right)^{n}, \label{eq:bv-n}
\end{eqnarray}
where $\tau$, $b$ and $n$ are varied. $k_{0}$ is the same pivot scale as Eq.~(\ref{eq:bv-2}). The quadratic power law model ($n=2$) is only a specific case of this model. $n=1$ and $n=3$ correspond to linear and cubic power law models respectively. Below, we name this model as the ``($\tau, b, n$) model''.
\end{enumerate}

\subsection{Likelihood analysis}

In Sec.~\ref{sec:data}, we explain that there could be an overall systematics of the amplitude of the correlation function data. In the likelihood function, we fit the correlation function model with different biases, both with and without the calibration parameter. For the likelihood without the calibration parameter, the $\chi^{2}$ function for each data set is
\begin{eqnarray}
\chi^{2}(\vec{\theta}) &=&  \sum^{N_{\rm bin}}_{i,j=1} \left(y^{\rm data}(r_{i})-y(r_{i},\vec{\theta}) \right) \left( C^{-1}\right)_{ij} \nonumber \\
& \times &\left(y^{\rm data}(r_{j})-y(r_{j},\vec{\theta}) \right),
\end{eqnarray}
where $\vec{\theta}=(\theta_{1},\theta_{2},...,\theta_{n})$ is the vector of free parameters. For the likelihood with a calibration parameter, the theoretical $y(r_{i},\vec{\theta})\rightarrow y(r_{i},\vec{\theta})+y_{0}$, where $y_{0}$ is the vertical calibration parameter. Because of the low value of $y_{0}$, we define $y_{9} \equiv y_{0} \times 10^{9}$ in the likelihood analysis.

The four data sets are produced from four foreground-cleaned maps, which are the results of different component separation algorithms. Therefore, once we obtain the individual $\chi^{2}$ function for each data set, we also use an averaged $\chi^{2}(\mathbf{\theta})$ to calculate the best-fit parameters, i.e.
\begin{eqnarray}
\chi^{2}_{\rm avg}=\left(\chi^{2}_{\sevem}+\chi^{2}_{\smica}+\chi^{2}_{\nilc}+\chi^{2}_{\commander} \right)/4.
\end{eqnarray}
Then the likelihood is $\mathcal{L} \sim \exp\left(-\chi^{2}/2 \right)$. We also marginalize to obtain the one-dimensional posterior distributions of each parameter.

\section{Results of likelihood analysis}
\label{sec:results}

\subsection{Likelihood without the calibration parameter}
We now analyse the results of the likelihood analysis without the parameter $y_{0}$. For the one parameter model ($\tau$ model), we apply a uniform prior over the range $\tau_{4}=[0,3]$ onto the likelihood function and plot its posterior distribution in Fig.~\ref{fig:tau_1p}. One can see that the four different data sets give consistent results of constraints. We list the values of the constraints in Table~\ref{tab:paras}, and compare the distribution of $\tau_{4}$ with the {\it Planck} posterior distribution in Fig.~\ref{fig:comp}. One can see that in Fig.~\ref{fig:comp} the blue dashed line favours a lower value of $\tau_{4}$, whereas the {\it Planck} result favours $\tau_{4} \simeq 1.39$~\citep{Planck16-unbound}.

We also plot the best-fit correlation function for the $\tau$ model as black solid line in Fig.~\ref{fig:deltaTv}.  One can see that the model predicts the strong correlations at small separation distance $r$ and becomes weaker at larger separation, but it does not recover the positive correlation at large $r$, which might be due to unaccounted systematics. In terms of goodness of fit, in Table~\ref{tab:chi2}, we list the minimal $\chi^{2}$ value for the four data sets we used, and their residual $\chi^{2}$, i.e. $\overline{\chi^{2}}_{\rm min}/N_{\rm dof}$. In the single-parameter model ($\tau_{4}$), $N_{\rm dof}=N_{\rm data}-N_{\rm para}=19$\footnote{In statistics, if $\overline{\chi^{2}}_{\rm min}/N_{\rm dof}$ is close to unity, it means that the model can provide a good fit to the data, and vice versa~\citep{Nesseris04}. This is because $\chi^{2}$ follows the $\chi^{2}$ distribution, so its expectation value of is $N_{\rm dof}=N_{\rm data}-N_{\rm samples}$~\citep{Riley06}.}. 

One can see that $\overline{\chi^{2}}_{\rm min}/N_{\rm dof}=2.91$ which is larger than unity, this indicates that the model does not provide a good fit to the data.

We then tried the quadratic power law model of the velocity bias model, i.e. Eq.~(\ref{eq:bv-2}). In this model, we vary both $\tau$ and $b$, and we multiply the flat prior on the likelihood as $\tau_{4}=[0,3]$ and $b=[-8,8]$. We plot the posterior distribution in Fig.~\ref{fig:fit_2p}. The panels (a) and (b) are for the marginalized distributions of $\tau_{4}$ and $b$ respectively, in which the black dashed, red dashed, green dashed and orange dashed lines are for \sevem, \smica, \nilc, and \commander data sets respectively, and the blue solid lines are for the averaged $\chi^{2}$. In panel (c), we plot the joint posterior distribution of $\tau_{4}$ and $b$ for the averaged $\chi^{2}$ function. One can see from panel (a) that the four data sets give consistent results on $\tau_{4}$ and the average $\chi^{2}$ give $\tau_{4}=0.51^{+0.14}_{-0.15}$ (Table~\ref{tab:paras}) which is a $\sim 3\sigma$ C.L. detection of non-zero optical depth. But the value of $\tau_{4}$ in this model is slightly higher than the single-parameter model (Table~\ref{tab:paras}). In panels (b) and (c) of Fig.~\ref{fig:fit_2p}, one can see that the posterior distribution of $b$ is skewed towards a negative value, and the best-fit value of $b$ for averaged $\chi^{2}$ is $b=-1.39^{+0.40}_{-0.57}$. The frequency that $b>0$ is only $4.47 \times 10^{-6}$, i.e. there is a very strong preference for the negative bias parameter. Comparing the residual $\chi^{2}$ for this model with the previous models in Table~\ref{tab:chi2}, one can see that $\overline{\chi^{2}}_{\rm min}/N_{\rm dof}$ drops from $2.91$ to $1.93$, but still much greater than unity. In Fig.~\ref{fig:deltaTv}, we plot the best-fit model prediction for the quadratic power law model as the blue dashed line. One can see that, since the velocity bias is included in the model, the shape of the function $y(r)$ becomes more consistent with the data sets. However, since $\overline{\chi^{2}}_{\rm min}/N_{\rm dof}$ is still greater than unity, we vary the power index $n$ to test whether it improves the fit.

We now vary the power index $n$ and run the varying power law model ($\tau,b,n$). We use the same flat prior for $\tau_{4}$ and $b$, and apply the flat prior over the range of $[-5,5]$ on $n$. In Fig.~\ref{fig:fit_3p}, we plot the marginalized posterior distribution for the three parameters on the first row (panels (a), (b) and (c)), and the marginalized two-dimensional constraints on the second row (panels (d), (e) and (f)). In panel (a), one can see that the marginalized distribution for $\tau_{4}$ is quite consistent between different data sets, and they are all very consistent with the constraint from the quadratic power law model (panel (a) in Fig.~\ref{fig:fit_2p}). The marginalized distribution of $b$, shown in panel (b) of Fig.~\ref{fig:fit_2p} slightly shifts to the less negative value. The averaged $\chi^{2}$ gives $b=-1.04^{+0.36}_{-0.53}$ ($68\%$ C.L.). The p-value ($P(b>0)$) is only $6.46 \times 10^{-6}$, which is a strong indication that the scale-dependence bias does exist for the peculiar velocity field at $k>0.1\,h\,{\rm Mpc}^{-1}$. More interesting is the marginalized posterior distribution of the spectral index $n$, shown in the panel (c). The best-fit value is $n=3.00^{+0.20}_{-0.49}$ which is $2\sigma$ C.L. away from the quadratic power law model ($n=2$), and $4\sigma$ C.L. away from linear power law model ($n=1$). It is at close to $4.5\sigma$ C.L. that the spectral index of the velocity bias is positive (panels (d), (e) and (f) in Fig.~\ref{fig:fit_3p}). In terms of p-value, $P(n<1)=3.32 \times 10^{-4}$, and $P(n<2)=0.059$. Therefore, spectral index with $k-$dependence of unity is strongly disfavoured, but it cannot exclude the quadratic power-law model.

Figure~\ref{fig:deltaTv} plots the best-fitting power law model as a purple solid line. One can see that the predicted correlation function is consistent with the amplitude and shape of the data sets at $r<80\,h^{-1}$Mpc, but at larger separation the prediction and the data sets are still inconsistent. The residual $\chi^{2}$ value in Table~\ref{tab:chi2} decreases by an additional $0.2$ for $\overline{\chi^{2}}_{\rm min}/N_{\rm dof}$. This indicates that the ($\tau,b,n$) model provides a slightly better fit to the data comparing to the other models, although the residual $\chi^{2}$ is still higher than unity.


\subsection{Likelihood with a calibration parameter}
We now analyse the results of the likelihood analysis with a varying parameter $y_{0}$. In Table~\ref{tab:paras} and Figs.~\ref{fig:scaling-2p}, \ref{fig:scaling-3p}, and \ref{fig:scaling-4p}, we vary the value of $y_{0}$ as a free parameter to control the total amplitude of the correlation function. We use this single parameter to take into account the subtraction effect of the aperture ring in between $8$\,arcmin and $8\times \sqrt{2}$\,arcmin, and other unaccounted systematics~\citep{Planck16-unbound}.

We first try the unbiased model with $\tau_{4}$ as a single parameter. In Fig.~\ref{fig:scaling-2p} and Table~\ref{tab:paras}, one can see that the value of $y_{9}$ ($y_{9}=y_{0}\times 10^{9}$) peaks at $5.26$ for the combined $\chi^{2}$, which is negligible. The constraints from different data sets are consistent with each other. However, by calculating $\overline{\chi^{2}}_{\rm min}/N_{\rm dof}$ (Table~\ref{tab:chi2}), we find that the value is comparable to the case of not including $y_{0}$ as a free parameter. So this model also cannot provide an excellent fit to the data (see the blue dashed line in the right panel of Fig.~\ref{fig:deltaTv}).

We further vary the value of $b$ to constrain the quadratic power-law model of the bias with a varying $y_{0}$ parameter. In Table~\ref{tab:chi2}, one can see that this model is by far the best model fitted to the data, with $\overline{\chi^{2}}/{N_{\rm dof}}=1.23$, close to unity. In Table~\ref{tab:paras} and Fig.~\ref{fig:scaling-3p}, one can see that the best-fit value of $\tau_{4}$ is $1.18 \pm 0.24$, which is close to what is obtained in~\citet{Planck16-unbound}. The distribution function of $b$ is negative, and the probability $P(b>0)$ is only $3.12 \times 10^{-8}$. The value of $y_{9}$ is $12.39^{+3.65}_{-3.66}$. In the right panel of Fig.~\ref{fig:deltaTv}, we can see that this model $b_v(k)=1+b(k/k_{0})^{2}$ provides the best match to the data, compared to all of the other models, with and without varying $y_{0}$ as a free parameter.

We further vary the power index $n$ as a free parameter in the model, and constrain the parameter sets ($\tau_{4}$, $b$, $n$, $y_{9}$). We show the results in Tables~\ref{tab:paras} and \ref{tab:chi2}, and Figs.~\ref{fig:deltaTv} \& \ref{fig:scaling-4p}. One can see that $\overline{\chi^{2}}_{\rm min}/N_{\rm dof}$ is not reduced for this case, and the fit to the data is not as good as the quadratic power law model. In panel (d) of Fig.~\ref{fig:scaling-4p}, the likelihood of $n$ broadly peaks at $n=2$, which indicates a preference for the quadratic power law model as the bias function.

In summary, we find that varying the $y_{0}$ parameter to allow the correlation function to shift vertically can accommodate the positive correlation of the data $\langle \Delta T v \rangle$ on scales of $r> 80\,h^{-1}$Mpc. The quadratic power-law model provides the best-fit results. The $\overline{\chi^{2}}_{\rm min}/N_{\rm dof}$ is reduced to $1.23$ in this case, and the model provides a good fit to data on all scales.

\section{Discussion}
\label{sec:discuss}

\begin{figure*}
\centerline{
\includegraphics[bb=0 0 743 464, width=3.4in]{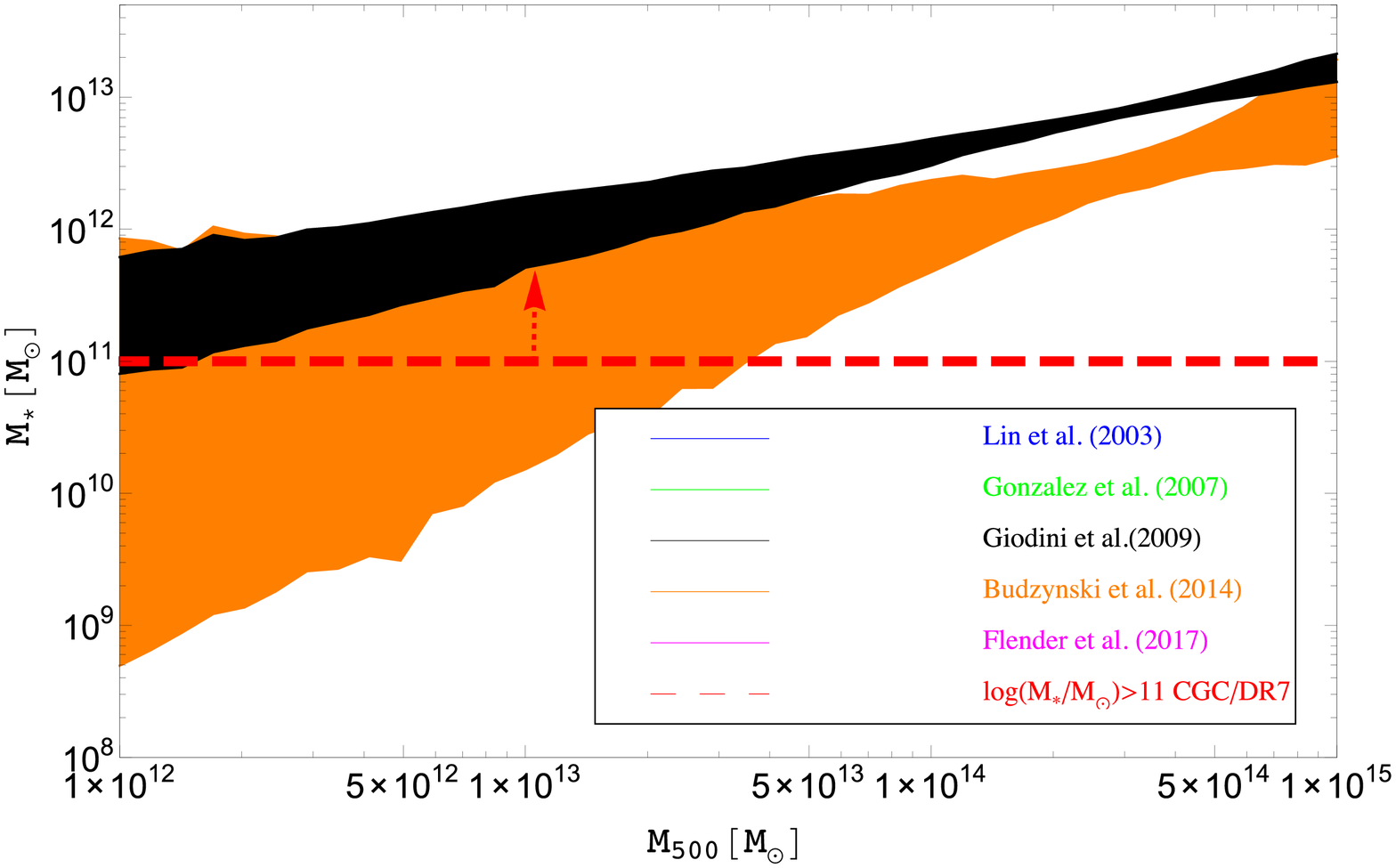}
\includegraphics[bb=0 -20 743 457, width=3.3in]{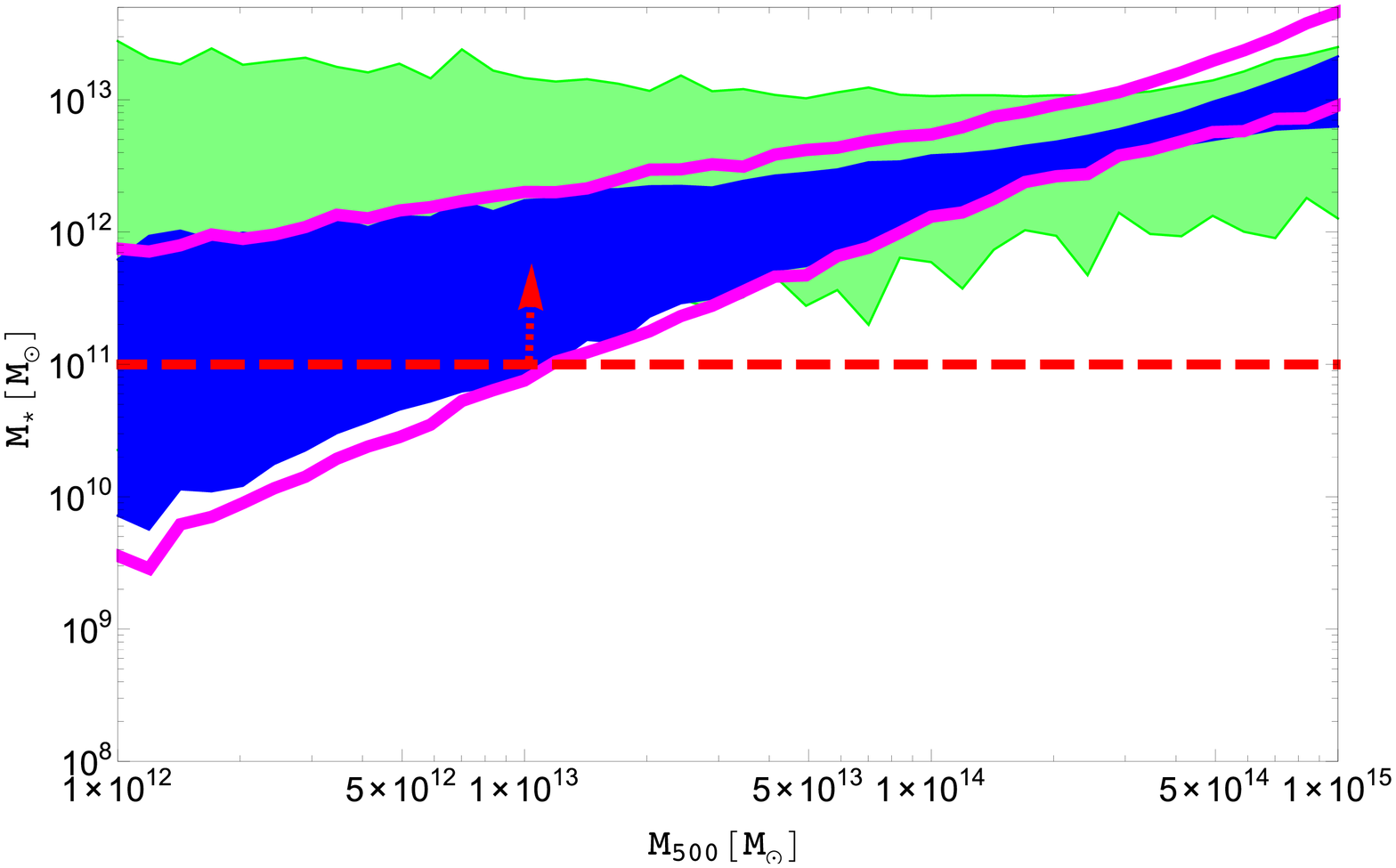}}
\caption{The $M_{\ast}$--$M_{500}$ relation reported in different literatures~(see also \citet{Flender17}). The red dashed line and upright arrow indicate that the CGC samples used in this study has $\log(M_{\ast}/\msun)>11$ and $M_{500}\simeq 10^{13}\msun$. The resulted region from the best-fit of~\citet{Flender17} is plotted as the red boundary lines, since its filled region almost overlaps with the blue region.} \label{fig:MsM500}
\end{figure*}

\begin{figure}
\centerline{
\includegraphics[width=3.4in]{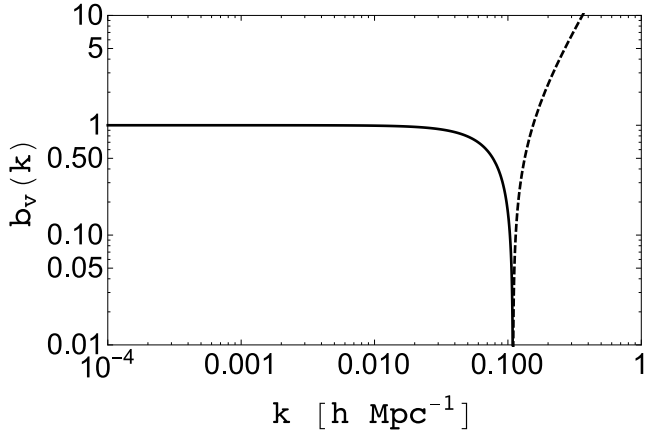}}
\caption{The best-fit velocity bias function, i.e. quadratic power-law model with calibration $y_{0}$, $b=-0.84$. The dashed line means the negative branch of the curve.} \label{fig:bv}
\end{figure}

\subsection{The optical depth}

In the left panel of Fig.~\ref{fig:comp}, we compare the one-dimensional posterior distribution of single-parameter model, and the marginalized distribution of the $\tau_{4}$ parameter of the quadratic power law and varying power law models, with and without varying the $y_{0}$ parameter, and the best-fit result of \citet{Planck16-unbound}. One can see that all of the models except the quadratic power-law model and the varying power-index model with free $y_{0}$ parameter prefer a lower value of $\tau_{4}$ compared to the {\it Planck} result (green line), with the center values differing at $\sim 2\sigma$ C.L. compared with {\it Planck}. However for quadratic power-law model (black solid line) and the varying power-index model (red solid line) with free $y_{0}$ parameter, the value of $\tau_{4}$ peaks at $\sim 1.18$ which is close to {\it Planck}'s peak value. The distribution are also quite close to each other. Note that~\citet{Planck16-unbound} derives the constraints on $\tau$ from a comparison between the cross-correlation data and an N-body simulation template for the kSZ temperature and velocity field correlation $\langle \Delta T v \rangle$.

The optical depth is a measurement of free electron density along the line of sight, and it is related to the mass, redshift and the density profiles of the samples. Different samples may have different values of the optical depth $\tau$. \citet{Birkinshaw99} used the ROSAT PSPC observational data and the spectral parameters of the X-ray emission found from ASCA data, to measure the emissivity of the intra-cluster gas of cluster CL 0016+16, and then derived the central density of the electron gas. This leads to the optical depth $\tau \sim 0.01\,h^{-1/2}_{100}$, which is higher than what we obtain\footnote{The reader should note that this value is derived from the X-ray observation, not directly measured from the kSZ observation.}.

The samples that we considered are not massive clusters, but central galaxies with stellar mass $\log(M_{\ast}/\msun)>11$~\citep{Planck16-unbound}. For the samples that we considered, the result of fitting the quadratic power-law model is consistent with the results of~\citet{Planck16-unbound}. We now want to investigate whether this is consistent with the general prediction of the optical depths of groups and clusters discussed in~\citet{Flender17}.~\citet{Flender17} provided a new model for the intracluster medium by taking into account the effects of star formation, feedback, non-thermal pressure, and gas cooling. The free parameters in this semi-analytic model were calibrated by using recent X-ray measurements of gas density profiles of clusters, and gas masses of groups and clusters. Figure~7 in~\citet{Flender17} plots the average optical depth as a function of $M_{500}$ for different redshift bins\footnote{$M_{500}$ is the total mass enclosed in a region with mean density equals to $500$ of the critical density of the Universe, i.e. $M_{500}=(4\pi/3)500\rho_{\rm crit}(z)R^{3}_{500}$.}. The optimal size $\theta=8\,$arcmin of our aperture photometry method is certainly beyond the virial radius of the CGC samples~\citep{Planck16-unbound}, so we refer to the $\Delta=200$ plot (panel (c)) in Fig.~7 of \citet{Flender17} to derive the mean value of our $M_{500}$ at $z\sim 0.1$. For $\tau \sim 1.18 \times 10^{-4}$, the corresponding $M_{500}$ is about $10^{13}\msun$. We now want to see whether this is consistent with the the range of stellar mass with which samples were selected ($\log(M_{\ast}/\msun)>11$). In~\citet{Flender17}, $M_{\ast}$ is related  to $M_{500}$ in the following fitting formula relates $M_{500}$ to $M_{\ast}$
\begin{eqnarray}
\left(\frac{M_{\ast}}{10^{12}\msun} \right)=3\times (100 f_{\ast})\left(\frac{M_{500}}{3\times 10^{14}\msun} \right)^{1-S_{\ast}}, \label{eq:M-ast}
\end{eqnarray}
where $f_{\ast}$ and $S_{\ast}$ are the amplitude and slope parameters, respectively\footnote{The original form in~\citet{Flender17} is given in $M_{\ast}/M_{500}$ and we convert it to this form.}. Table~\ref{tab:Ms} shows the different values of $f_{\ast}$ and $S_{\ast}$ reported in various works. The quoted error-bars indicate the $1\sigma$ confidence levels. Since both of the parameters are varying, we generate $10^{4}$ random numbers from each of the Gaussian distributions of $f_{\ast}$ and $S_{\ast}$, and then found the maximum and minimum values of $M_{\ast}$ for given $M_{500}$. We report our boundary lines in Fig.~\ref{fig:MsM500}. The filled regions of blue, green, black and orange colours are the previous results from~\citet{Lin03}, \citet{Gonzalez07}, \citet{Giodini09} and \citet{Budzynski14} respectively. The result of ~\citet{Flender17} is plotted as the boundary lines with magenta colour since it almost overlaps with ~\citet{Lin03} (blue region). One can see that, for $M_{500}\simeq 10^{13}\msun$, the corresponding $M_{\ast}$ predicted from the current fitting formula has a large variance. Taking the~\citet{Flender17} and \citet{Lin03} results as examples: The $M_{\ast}$ could be in the range of $10^{11}\msun$--$2 \times 10^{12}\msun$, which is consistent with our sample selection criterion~\citep{Planck16-unbound}. Note that Table~B.1 in~\citet{Planck13-XI} and Table~1 in~\citet{Anderson15} also show that galaxies with stellar mass $\log(M_{\ast}/\msun)$ corresponds to host halos with mass $\log(M_{\rm h}/\msun)\simeq 13$. The~\citet{Budzynski14} and \citet{Gonzalez07} results have larger variances: for $M_{500}\simeq 10^{13}\msun$, $M_{\ast}$ is in the range of [$1.5 \times 10^{10}$,$1.1\times 10^{12}$]$\msun$ for \citet{Budzynski14} and [$9.7 \times 10^{10}$,$1.4\times 10^{13}$]$\msun$ for \citet{Gonzalez07} respectively. These ranges clearly cover the range of the CGC sample selection ($M_{\ast}>10^{11}\msun$). Only the~\citet{Giodini09} prefers a slightly a higher range of $M_{\ast}$, i.e. [$5.1 \times 10^{11}$, $1.8 \times 10^{12}$]$\msun$ for $M_{500}\simeq 10^{13}\msun$ which is sightly higher than the expected range of $M_{\ast}$.

To summarize, the $M_{\ast}$--$M_{500}$--$\tau$ relation for our selected samples and final value of $\tau$
\begin{eqnarray}
\log\left(\frac{M_{\ast}}{\msun} \right)>11 \,\,\Rightarrow M_{500}\simeq 10^{13}\msun \,\, \Rightarrow \tau =1.18 \times 10^{-4}
\end{eqnarray}
is consistent with the fitting formula derived from observations \citep{Lin03,Gonzalez07,Budzynski14,Flender17}.

\begin{table}
\begin{centering}
\begin{tabular}{|c|c|c|}
\hline
\noalign{\vskip 1pt}
Reference & $100f_{\ast}$ & $S_{\ast}$ \\
\hline
\citet{Lin03} & $1.64^{+0.10}_{-0.09}$ & $0.26 \pm 0.09$ \\ \hline
\citet{Gonzalez07} & $2.02 \pm 0.37$ & $0.64 \pm 0.13$  \\ \hline
\citet{Giodini09} & $2.58 \pm 0.05$ & $0.37 \pm 0.04$ \\ \hline
\citet{Budzynski14} & $0.912 \pm 0.06$ & $0.11 \pm 0.14$ \\ \hline
\citet{Flender17} & $2.6 \pm 0.3$ & $0.12 \pm 0.1$ \\ \hline
\end{tabular}
\caption{Values of $f_{\ast}$ and $S_{\ast}$ for Eq.~(\ref{eq:M-ast}) reported in different literatures (see also~\citet{Flender17}). The $M_{\ast}$--$M_{500}$ relation for these five models are plotted in Fig.~\ref{fig:MsM500}.} \label{tab:Ms}
\end{centering}
\end{table}

\subsection{The velocity bias}

Figure~\ref{fig:bv} plots the best-fit function of the velocity bias. As one can see, on very large-scales, the velocity bias is almost unity for $k \leq 0.1 h\,$Mpc$^{-1}$, but because of the negative sign of the best-fit $b$ value, the bias tends to drop below unity at larger $k$. However, this does not affect the correlation function very much, because at very large $k$, the angular averaged function $F_{r}(k)$ (Eq.~(\ref{eq:Fr_k})) becomes oscillatory and approaches zero at $k \rightarrow 1\,h\,{\rm Mpc}^{-1}$, and hence the combined effects cancel at very large $k$. Therefore, the real improvement of the constraint comes from the intermediate scale close to the pivot scale, i.e. $k=0.01$--$1\,h\,$Mpc$^{-1}$. Note that the CGC samples selected by \citet{Planck16-unbound} have $\log(M_{\star}/M_{\odot})>11$. According to Table B.1. in \citet{Planck13-XI} and Table~1 in~\citet{Anderson15}, galaxies with this range of stellar mass normally reside in halos with mass $\log(M_{\rm h}/M_{\odot}) \simeq  13$.

Comparing with the simulation results in \citet{Zhang15} and \citet{Zheng15a,Zheng15b}, with halo mass in the range of $10^{12}$--$10^{13}\,h^{-1}\,M_{\odot}$, $b_{v}$ is close to unity within $2\%$ model uncertainty at $k\leq 0.1\,h\,{\rm Mpc}^{-1}$ in the redshift range of $z=0$--$2$, while at $k \geq 0.1\,h\,{\rm Mpc}^{-1}$, $b_{v}$ drops below unity. So our results are broadly consistent with the findings in \citet{Zhang15} and~\citet{Zheng15a,Zheng15b}, although correspond to a slightly larger range of halo mass\footnote{According to eq.~(3) in~\citet{Zheng15b}, their defined velocity bias is the square root of the velocity bias we defined here (Eq.~(\ref{eq:bv-def})).}. Theoretically, this negative sign of $b$ indicates that on mildly non-linear scales (large $k-$values), the peak velocities can anti-correlate with the underlying dark matter distribution. This is consistent with the calculation shown in \citet{Baldauf15} and \citet{Chan15}, i.e. the difference between the peak velocity and smoothed velocity is a negative term proportional to the gradient of dark matter density.

In addition, the results of our likelihood analysis show that the quadratic power law model $n=2$ provides the best-fit to the data, which is consistent with the prediction of velocity bias in the peak theory of density fluctuations~\citep{Desjacques08,Desjacques10,Baldauf15,Chan15}. As shown in~\citet{Desjacques08} and \citet{Desjacques10}, the large-scale limit of the peak 2-point correlation and mean pairwise velocity can be thought as arising from the relation $\delta_{\rm pk}(\mathbf{x})=b_{10}\delta_{\rm m}(\mathbf{x})-b_{01}\nabla^{2}\delta_{\rm m}(\mathbf{x})$, $\mathbf{v}_{\rm pk}(\mathbf{x})=\mathbf{v}_{\rm m}(\mathbf{x})-R^{2}_{v}\nabla\delta_{\rm m}(\mathbf{x})$, where $R_{v}$ is the characteristic scale of peak velocity bias. One can see that in the continuous approach of peak background-split theory, the peak velocity field is unbiased with respect to the matter velocity field $\delta_{\rm m}(\mathbf{x})$, but it receives a contribution from the first derivative of the density $\nabla \delta_{\rm m}(\mathbf{x})$. This means that the peak pairwise velocity, or mean streaming, is obtained from the statistics of the (proper) matter velocity field $\mathbf{v}_{\rm m}(\mathbf{x})$, but the latter has to be evaluated at those maxima of the density field. Therefore $\mathbf{v}_{\rm pk}$ receives a contribution to the density gradient. \citet{Desjacques08} showed that, provided that the peak velocity field receives a contribution from the gradient of the density field, the peak theory is consistent with the nonlinear local biasing relation inferred from the 2-point correlation of density maxima. Physically, this indicates that on the small scales ($k \gg 0.1\,h\,{\rm Mpc}^{-1}$) the gas velocity field is anti-correlated with the matter density field and the gas tends to be pushed out of the dark matter halos. We plan to investigate this effect further by using hydrodynamic simulations.

Thus, as shown in~\citet{Desjacques08}, \citet{Desjacques10}, \citet{Baldauf15} and \citet{Chan15}, the peak density is related to the smoothed density via $\delta_{\rm pk}(\mathbf{x})=b_{10}\delta_{\rm m}(\mathbf{x})-b_{01}\nabla^{2}\delta_{\rm m}(\mathbf{x})$, which in Fourier space is $\mathbf{v}_{\rm pk} \simeq (1-b_{01}k^{2})\mathbf{v}_{\rm m}$. Thus our fitting results observationally verify their prediction. If we apply the Poisson equation to the $\delta_{\rm pk}$ equation, we find that in order to reconstruct the gravitational potential at the peak position, one has to include the $k-$dependent bias as well, i.e.
\begin{eqnarray}
\phi_{\rm pk}(\mathbf{x})=(b_{10}+b_{01}k^{2})\phi(\mathbf{x}),
\end{eqnarray}
where $\phi(\mathbf{x})$ is the gravitational potential for the total matter field.

%

\section{Conclusion and Prospects}
\label{sec:conclusion}

In this paper, we probed the optical depth of the Central Galaxy Catalogue (CGC) of SDSS DR7 as well as the scale-dependence of the velocity bias by using the cross-correlation data of kSZ--velocity field. We first present the kSZ--velocity cross-correlation data we use in this work, derived from {\it Planck} \sevem, \smica, \nilc and \commander\, foreground-cleaned maps and the CGC data. Then we presented the theoretical calculation of the $\langle (\Delta T/T)(\mathbf{v}\cdot \hat{\mathbf{n}}/\sigma_{v}) \rangle$ from the linear perturbation theory. In addition, we reviewed three models of velocity bias, namely a unity bias, a quadratic power-law model, and a varying power-law model with one, two and three free parameters respectively. We further fitted the three models individually with the correlation data, with and without the free $y_{0}$ parameter that takes into account the possible subtraction effect on large scales of correlation ($r>100\,h^{-1}$Mpc).

We calculated the $\chi^{2}$ value for each model separately, and then combined them to calculate the average $\chi^{2}$. Our results showed that the quadratic power-law model $b_v(k)=1+b(k/k_0)^{2}$ with varying $y_{0}$ parameter provides the best-fit model to the data, and the constrained parameters are $b=-0.84^{+0.16}_{-0.20}$, $\tau=(1.18 \pm 0.24)\times 10^{-4}$, and $y_{0}=(12.39^{+3.65}_{-3.66}) \times 10^{-9}$. This strongly indicates that the bias $b$ is negative, with the probability $P(b>0)=3.12 \times 10^{-8}$. The reduced $\chi^{2}$ value is equal to $1.23$. By comparing the correlation function of the model with the data, one can find that the model matches the data at all scales. We compared the results of the distribution of $\tau_{4}$ with {\it Planck} results, and show that the result found from this quadratic power-law model is the closest to {\it Planck} data. We found that this value of $\tau$ corresponds to the halo mass $M_{500}\simeq 10^{13}\msun$, which is consistent with the selection criterion $M_{\ast}>10^{11}\msun$ of the CGC samples, given the fitting formula of $M_{\ast}$--$M_{500}$ derived by~\citet{Flender17}. The functional form of $b_{v}\simeq 1-(k/k_{0})^{2}$ is consistent with the peak-background split theory~\citep{Desjacques08,Desjacques10}, i.e. that the peak density linearly follows the density of dark matter field, but receives a gradient contribution from $\delta_{\rm m}$. Our result verifies the conclusion of~\citet{Desjacques08} and \citet{Desjacques10}.

In the future, there will be more precise data to improve such constraints. The SDSS-IV project started to operate in 2014, and it will provide a much larger spectroscopic catalogue, deep into $0.2 \lesssim z \lesssim 0.75$ regime~\citep{Wang17}. By cross-correlating SDSS-IV data with {\it Planck} map, one can also constrain the redshift evolution of velocity bias. In summary, the cross-correlation between kSZ temperature fluctuations and the linear velocity field provides a direct and powerful tool to constrain peculiar velocity bias, and improve our understanding of the formation of the large-scale structure of the Universe.

\vskip 0.1 truein

\noindent \textbf{Acknowledgments:} We are grateful for Carlos Hern{\'a}ndez-Monteagudo for sharing the kSZ--velocity field cross-correlation data and many helpful discussions. In addition, we also thank Enzo Branchini, Anthony Challinor, Matt Hilton, Hideki Tanimura and the anonymous referee for helpful discussions and suggestions. Y.Z.M. acknowledges the support by National Research Foundation of South Africa (no. 105925). P.H. acknowledges the support by the National Science Foundation of China (No. 11273013).

\appendix


\begin{thebibliography}{}

\bibitem[Anderson et al.(2015)]{Anderson15} Anderson, M.~E., Gaspari, M., White, S.~D.~M., Wang, W., \& Dai, X.\ 2015, \mnras, 449, 3806

\bibitem[Baldauf et al.(2015)]{Baldauf15} Baldauf, T., Desjacques, V., \& Seljak, U.\ 2015, \prd, 92, 123507

\bibitem[Blanton et al.(2005)]{Blanton05} Blanton, M.~R., Schlegel, D.~J., Strauss, M.~A., et al.\ 2005, \aj, 129, 2562

\bibitem[Biagetti et al.(2014)]{Biagetti14} Biagetti, M., Desjacques, V., Kehagias, A., \& Riotto, A.\ 2014, \prd, 90, 103529

\bibitem[Birkinshaw(1999)]{Birkinshaw99} Birkinshaw, M.\ 1999, \physrep, 310, 97


\bibitem[Budzynski et al.(2014)]{Budzynski14} Budzynski, J.~M., Koposov, S.~E., McCarthy, I.~G., \& Belokurov, V.\ 2014, \mnras, 437, 1362


\bibitem[Chan(2015)]{Chan15} Chan, K.~C.\ 2015, \prd, 92, 123525

\bibitem[De Bernardis et al.(2017)]{Bernardis16} De Bernardis, F., Aiola, S., Vavagiakis, E.~M., et al.\ 2017, \jcap, 3, 008

\bibitem[Desjacques(2008)]{Desjacques08} Desjacques, V.\ 2008, \prd, 78, 103503

\bibitem[Desjacques \& Sheth(2010)]{Desjacques10} Desjacques, V., \& Sheth, R.~K.\ 2010, \prd, 81, 023526


\bibitem[Elia et al.(2012)]{Elia12} Elia, A., Ludlow, A.~D., \& Porciani, C.\ 2012, \mnras, 421, 3472

\bibitem[Ferraro et al.(2016)]{Ferraro16} Ferraro, S., Hill, J.~C., Battaglia, N., Liu, J., \& Spergel, D.~N.\ 2016, \prd, 94, 123526

\bibitem[Flender et al.(2017)]{Flender17} Flender, S., Nagai, D., \& McDonald, M.\ 2017, \apj, 837, 124

\bibitem[Giodini et al.(2009)]{Giodini09} Giodini, S., Pierini, D., Finoguenov, A., et al.\ 2009, \apj, 703, 982

\bibitem[Gonzalez et al.(2007)]{Gonzalez07} Gonzalez, A.~H., Zaritsky, D., \& Zabludoff, A.~I.\ 2007, \apj, 666, 147


\bibitem[Guo et al.(2015a)]{Guo15a} Guo, H., Zheng, Z., Zehavi, I., et al.\ 2015, \mnras, 453, 4368

\bibitem[Guo et al.(2015b)]{Guo15b} Guo, H., Zheng, Z., Zehavi, I., et al.\ 2015, \mnras, 446, 578


\bibitem[Hand et al.(2012)]{Hand12} Hand, N., Addison, G.~E., Aubourg, E., et al.\ 2012, Physical Review Letters, 109, 041101

\bibitem[Hern{\'a}ndez-Monteagudo et al.(2015)]{Carlos} Hern{\'a}ndez-Monteagudo, C., Ma, Y.-Z., Kitaura, F.~S., et al.\ 2015, Physical Review Letters, 115, 191301

\bibitem[Hill et al.(2016)]{Hill16} Hill, J.~C., Ferraro, S., Battaglia, N., Liu, J., \& Spergel, D.~N.\ 2016, Physical Review Letters, 117, 051301

\bibitem[Lin et al.(2003)]{Lin03} Lin, Y.-T., Mohr, J.~J., \& Stanford, S.~A.\ 2003, \apj, 591, 749


\bibitem[Ma et al.(2011)]{Ma11} Ma, Y.-Z., Gordon, C., \& Feldman, H.~A.\ 2011, \prd, 83, 103002

\bibitem[Nesseris \& Perivolaropoulos(2004)]{Nesseris04} Nesseris, S., \& Perivolaropoulos, L.\ 2004, \prd, 70, 043531

\bibitem[Nozawa et al.(1998)]{Nozawa98} Nozawa, S., Itoh, N., \& Kohyama, Y.\ 1998, \apj, 508, 17

\bibitem[Nozawa \& Kohyama(2015)]{Nozawa15} Nozawa, S., \& Kohyama, Y.\ 2015, Astroparticle Physics, 62, 30

\bibitem[Peebles(1993)]{Peebles93} Peebles, P.~J.~E.\ 1993, Principles of Physical Cosmology by P.J.E.~Peebles.~Princeton University Press, 1993.~ISBN: 978-0-691-01933-8,

\bibitem[Percival \& Sch{\"a}fer(2008)]{Percival08} Percival, W.~J., \& Sch{\"a}fer, B.~M.\ 2008, \mnras, 385, L78

\bibitem[Planck Collaboration et al.(2013)]{Planck13-XI} Planck Collaboration, Ade, P.~A.~R., Aghanim, N., et al.\ 2013, \aap, 557, A52

\bibitem[Planck Collaboration et al.(2016a)]{Planck16-unbound} Planck Collaboration, Ade, P.~A.~R., Aghanim, N., et al.\ 2016a, \aap, 586, A140

\bibitem[Planck Collaboration et al.(2016b)]{Planck15-X} Planck Collaboration, Adam, R., Ade, P.~A.~R., et al.\ 2016b, \aap, 594, A9

\bibitem[Planck Collaboration et al.(2016c)]{Planck15-para} Planck Collaboration, Ade, P.~A.~R., Aghanim, N., et al.\ 2016c, \aap, 594, A13

\bibitem[Riley, Hobson, \& Bence (2006)]{Riley06} Riley~K.F., Hobson~M.P., \& Bence~S.J.~2006, Mathematical Methods for Physics and Engineering,~Cambridge University Press, 2006.

\bibitem[Schaan et al.(2016)]{Schaan16} Schaan, E., Ferraro, S., Vargas-Maga{\~n}a, M., et al.\ 2016, \prd, 93, 082002

\bibitem[Soergel et al.(2016)]{Soergel} Soergel, B., Flender, S., Story, K.~T., et al.\ 2016, \mnras, 461, 3172


\bibitem[Sunyaev \& Zeldovich(1972)]{Sunyaev72} Sunyaev, R.~A., \& Zeldovich, Y.~B.\ 1972, Comments on Astrophysics and Space Physics, 4, 173

\bibitem[Sunyaev \& Zeldovich(1980)]{Sunyaev80} Sunyaev, R.~A., \& Zeldovich, I.~B.\ 1980, \mnras, 190, 413

\bibitem[Wang et al.(2017)]{Wang17} Wang, Y., Zhao, G.-B., Chuang, C.-H., et al.\ 2017, \mnras, 469, 3762

\bibitem[Zhang et al.(2015)]{Zhang15} Zhang, P., Zheng, Y., \& Jing, Y.\ 2015, \prd, 91, 043522

\bibitem[Zheng et al.(2015a)]{Zheng15a} Zheng, Y., Zhang, P., \& Jing, Y.\ 2015, \prd, 91, 043523

\bibitem[Zheng et al.(2015b)]{Zheng15b} Zheng, Y., Zhang, P., \& Jing, Y.\ 2015, \prd, 91, 123512





































\end{thebibliography}
\end{document}